\documentclass[%
 aip,nofootinbib,
 jcp,%aps,
 amsmath,amssymb,
a4paper,11pt]{revtex4-1}

\usepackage{graphicx}% Include figure files
\usepackage{dcolumn}% Align table columns on decimal point
\usepackage{bm}% bold math
\begin{document}

%\preprint{AIP/123-QED}

\title{Proofreading of DNA Polymerase: a new kinetic model with higher-order terminal effects}
%\thanks{Footnote to title of article.}

\author{Yong-Shun Song}
 \affiliation{
School of Physical Science, University of Chinese Academy of Sciences
}%
\author{Yao-Gen Shu}%
\affiliation{
Institute of Theoretical Physics, Chinese Academy of Sciences
}%

\author{Xin Zhou}
\affiliation{
School of Physical Science, University of Chinese Academy of Sciences
}%

\author{Zhong-Can Ou-Yang}%
\affiliation{
Institute of Theoretical Physics, Chinese Academy of Sciences
}%

\author{Ming Li}
\email{liming@ucas.ac.cn}
\affiliation{
School of Physical Science, University of Chinese Academy of Sciences
}%

\date{\today}

\begin{abstract}
The fidelity of DNA replication by DNA polymerase (DNAP) has long been an important issue in biology. While numerous experiments have revealed details of the molecular structure and working mechanism of DNAP which consists of both a polymerase site and an exonuclease (proofreading) site, there were quite few theoretical studies on the fidelity issue.
The first model which explicitly considered both sites was proposed in 1970s' and the basic idea was widely accepted by later models.
However, all these models did not systematically and rigorously investigate the dominant factor on DNAP fidelity, i.e, the higher-order terminal effects through which the polymerization pathway and the proofreading pathway coordinate to achieve high fidelity.
In this paper, we propose a new and comprehensive kinetic model of DNAP based on some recent experimental observations, which includes previous models as special cases.
We present a rigorous and unified treatment of the corresponding steady-state kinetic equations of any-order terminal effects, and derive analytical expressions for fidelity in terms of kinetic parameters under bio-relevant conditions.
These expressions offer new insights on how the the higher-order terminal effects contribute substantially to the fidelity in an order-by-order way, and also show that the polymerization-and-proofreading mechanism is dominated only by very few key parameters.
We then apply these results to calculate the fidelity of some real DNAPs, which are in good agreements with previous intuitive estimates given by experimentalists.

\end{abstract}

%Use showkeys class option if keyword
\maketitle

\section{Introduction}
Since the Watson-Crick base-pairing rules of double-strand DNA was established\cite{watson1953molecular}, template-directed DNA replication became an important issue both in basic researches and application studies (e.g., Polymerase Chain Reaction ) in biology. The match between the incoming nucleotide dNTP and the template (i.e., the canonical Watson-Crick base pairing A-T and G-C) in the replication process plays a central role for any organism to maintain its genome stability, whereas mismatch (non-canonical base pairing like A-C) may introduce harmful genetic variations into the genome, and thus the error rate of replication must be kept very low. In living cells, the replication fidelity is controlled mainly by DNA polymerase (DNAP)\cite{lehman1958enzymatic} which catalyzes the template-directed DNA synthesis, and the fidelity of DNAP has been intensively studied since its discovery in 1950s'.\cite{kunkel2000dna}.

\begin{figure}
\includegraphics[width=.8\textwidth]{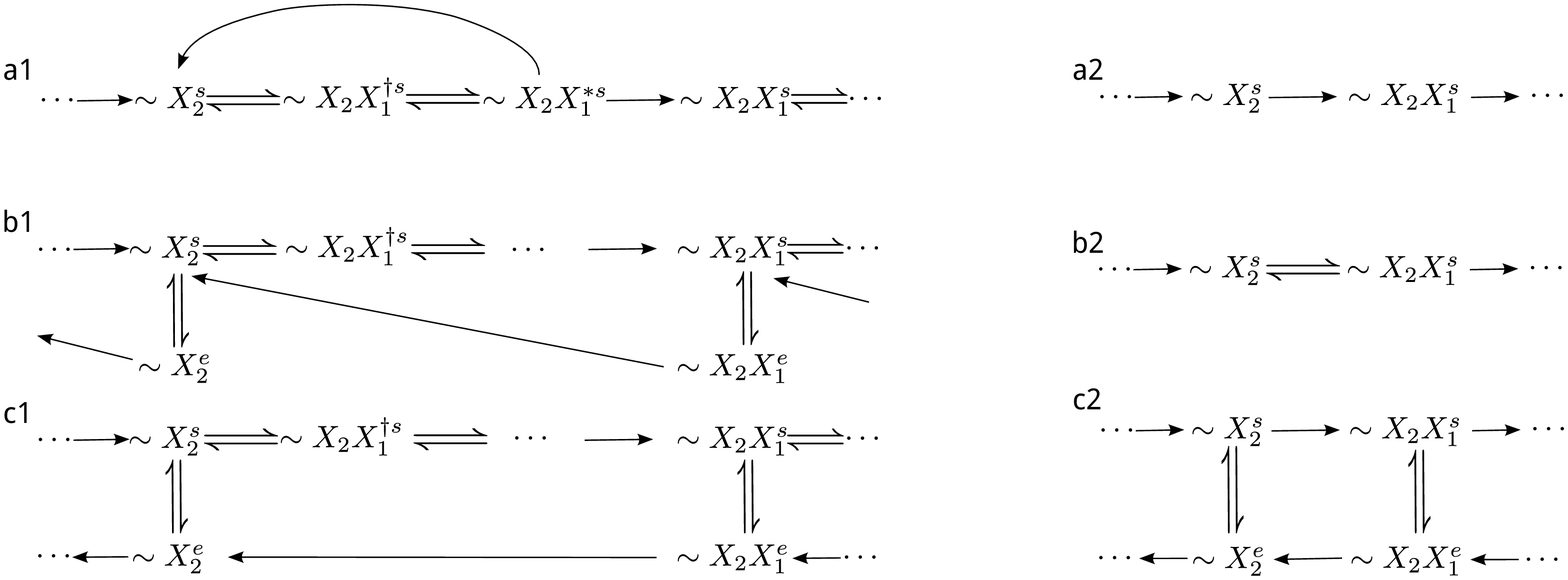}
\caption{Three different proofreading mechanisms of DNAP. For simplification, matched or mismatched dNTP is represented by $A$ or $B$ respectively throughout this paper. $X_i(i=\cdots, 1,2,\cdots)$ denotes either $A$ or $B$. The superscript $s$ or $e$ means that the primer terminus is in the polymerase (i.e., synthetic) site or the exonuclease site, respectively. $\sim X_2X_1^{\dagger s}$ denotes the state dNTP binding to DNAP before DNAP undergoes further conformation change. (a1) The original kinetic proofreading mechanism. $\sim X_2X_1^{\ast s}$ represents one or more high-energy intermediate states which dissociate much faster for $B$ than for $A$. This proofreading occurs before the nucleotide is covalently incorporated into the primer. (b1) Brief sketch of the Galas-Branscomb model. (c1) An alternative exonuclease proofreading mechanism proposed in this paper. Considering only the exact calculation of the fidelity, one can simplify these schemes under steady-state conditions, i.e., (a1) can be simplified as minimal scheme (a2), (b1) simplified as (b2), and (c1) simplified as (c2).$^{1}$}
\label{scheme_compare}
\end{figure}

\begin{figure}
\includegraphics[width=.8\textwidth]{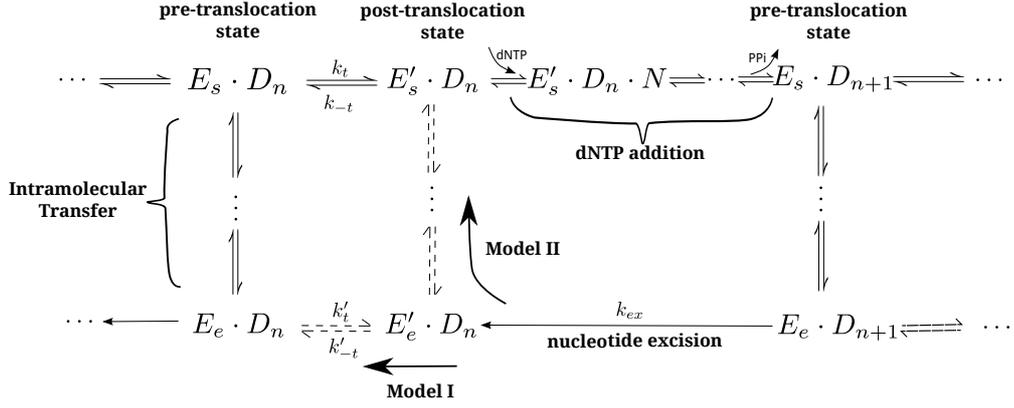}
\caption{The reaction scheme for DNAP. $E$: DNA polymerase; $D_n$: the state of the primer, $n$ being the length of the primer; $N$: dNTP. $E_s\cdot D_{n}$ and $E_s^{\prime}\cdot D_{n}$: the `polymerase type' complex when DNAP is in the pre-translocation state and post-translocation state, respectively. $E_e\cdot D_{n}$ and $E_e^{\prime}\cdot D_{n}$: the `exonuclease type' complex when DNAP is in the pre-translocation state and post-translocation state, respectively. A free dNTP can bind to DNAP when the complex is at the post-translocation state $E_s^\prime\cdot D_n$. When the dNTP is incorporated into the primer, the complex will return to the pre-translocation state $E_s\cdot D_{n+1}$. The primer terminus may be unzipped from the duplex and transferred to the exonuclease site. Model I and Model II indicate two possible pathways after the nucleotide excision in the exonuclease site.}
\label{scheme_full}
\end{figure}

Pioneering theoretical studies on this issue were done by J.Hopfield\cite{hopfield1974kinetic} and J.Ninio\cite{ninio1975kinetic}. Regarding DNA replication approximately as a binary copolymerization process of matched nucleotides (denoted as A for convenience in the present paper) and mismatched nucleotides (denoted as B), they proposed independently the so-called kinetic proofreading mechanism which correctly points out that the replication fidelity is not determined thermodynamically by the free energy difference, but kinetically by the incorporation rate difference,
between the match and the mismatch. This model, however, assumed that the proofreading occurs before nucleotide incorporation is accomplished (as illustrated in FIG.~\ref{scheme_compare}(a1)), which is not the case of real DNAPs.
Structural and functional studies show that DNAP often has two parts.
The basic part of all DNAPs is a synthetic domain (i.e., polymerase ) which binds the incoming dNTP and catalyzes its incorporation into the nascent ssDNA strand (called as primer below for convenience).
Proofreading is performed by a second domain (i.e, exonuclease) which is not a necessary part of DNAP.
This domain may much likely excise the just-incorporated mismatched nucleotide, once the mismatched terminus is transferred from the polymerase site into the exonuclease site by thermal fluctuation.
The first model that explicitly invokes the exonuclease, referred to as Galas-Branscomb model (FIG.~\ref{scheme_compare}(b1)), was proposed by Galas et al. \cite{galas1978enzymatic} and revisited by many other groups \cite{clayton1979error,johnson1993conformational,goodman1997hydrogen,goodman1998dna}.
Many experimental studies gave consistent results to this model\cite{fersht1979fidelity,patel1991pre,Cline1996PCR}.
Recently, improved experimental techniques revealed more details of the synthesizing and proofreading processes\cite{wuite2000single,tsai2006new}, and several detailed kinetic models have been proposed\cite{tsai2006new,xie2009possible,sharma2012error}.
However, all these models are based on the original simple Galas-Branscomb model and many important details such as higher-order neighbor effects of the primer terminus are not considered systematically\cite{sharma2012error} (see later sections).
In particular, recent experimental works on phi29 DNA polymerase \cite{lieberman2013kinetic,lieberman2014kinetic} revealed more details about the working mechanism of DNAP, highlighting the importance of the forward and backward translocation steps which were absent from the Galas-Branscomb models. Considering this point, as well as many other structural\cite{ollis1985structure,berman2007structures,doublie1998crystal,kamtekar2004insights,wang1997crystal} and kinetic\cite{patel1991pre,donlin1991kinetic,johnson1993conformational,lieberman2013kinetic,lieberman2014kinetic} experimental results, we propose a comprehensive reaction scheme of DNAPs as shown in FIG.~\ref{scheme_full}.

\footnotetext{In Section \ref{sec:fidelity}, the replication fidelity is defined as the ratio between the steady-state flux $J_A$ and $J_B$. In order to calculate such steady-state flux-flux ratios, one can map the original schemes to much simplified versions. For instance, any multistep pathway without branches can be mapped to a single-step pathway. Thus one obtains the much simplified schemes (a2), (b2) and (c2). On the other hand, in many kinetic assays of the DNAP reactions, multiple steps in the same pathway cannot be identified individually. In such cases, minimal schemes like (a2), (b2) and (c2) are directly used to analyze the experimental data.}

There are several key features of this scheme. First, the template-primer duplex binds to DNAP and forms two types of complexes. In the `polymerase type', the 3' terminus of the primer is located at the polymerase site. In the `exonuclease type', the primer terminus is unzipped from the duplex and transferred to the exonuclease site.
For the `polymerase type' complexes, two substates were experimentally observed \cite{lieberman2013kinetic,lieberman2014kinetic}.
One is the pre-translocation state of DNAP in which the dNTP binding site is occupied by the primer terminus.
The other is the post-translocation state in which the DNAP translocates forward (relative to the template) to expose the binding site to the next dNTP. DNAP can rapidly switch between these two states.
Correspondingly, one can assume two substates of DNAP in the `exonuclease type' complexes, though there are not sufficient experimental evidences.
One is the pre-translocation state in which the exonuclease site is occupied when the primer terminus is transferred from the polymerase site.
The other is the post-translocation state in which the exonuclease site is exposed after the nucleotide excision while the newly-formed primer terminus does not return to the polymerase site.

Second, once the incoming dNTP is incorporated into the primer, the DNAP can either translocate forward to the post-translocation state and bind a new dNTP in the polymerase site, or it pauses at the pre-translocation state and the primer terminus is unzipped from the duplex and transferred to the exonuclease site (the terminus can switch between the two sites without being excised\cite{lieberman2014kinetic}).
The large distance about $30-40$ \AA \cite{ollis1985structure,berman2007structures,doublie1998crystal,kamtekar2004insights,wang1997crystal} between the two sites implies that more than one nucleotides of the primer terminus must be unzipped, and thus the stability of the entire terminal region may put an impact on the unzipping probability of the primer terminus. Such neighbor effects, as well as other types of neighbor effects, can be very significant for the replication fidelity and should be taken account of in the kinetic models (details see later sections).

\begin{figure}
\includegraphics[width=.5\textwidth]{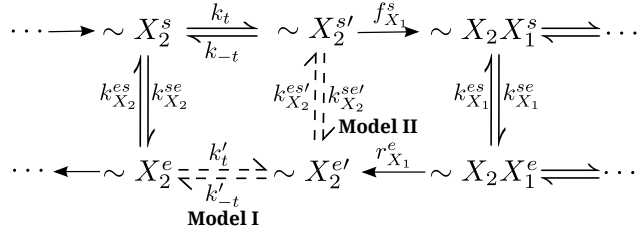}
\caption{The simplified reaction scheme. $X^s$, $X^e$ (or $X^{s\prime}$, $X^{e\prime}$): pre-translocation (or post-translocation) state of DNAP when the primer terminus is in the synthetic(s) site or the exonuclease(e) site respectively. When the primer terminus is in the exonuclease site, one does not need to distinguish between $\sim A^e(\sim B^e)$. However, it's still convenient to use $\sim A^e(\sim B^e)$ to denote the immediate state when the terminus switches back to the polymerase site.
By setting all the excision rates equal to $r^e$, we obtain the models for real DNAPs. Under the steady-state conditions, the effective rate of dNTP addition can be expressed as $f_{X_1}^s=k_{p}[X_1]$, where $k_{p}$ is an effective quasi-first-order rate constant, $[X_1]$ is the concentration of the incoming dNTP (to calculate the intrinsic fidelity, one often sets $[A]=[B]$). All other kinetic parameters in this figure are effective parameters which are combinations of the original rate constants in FIG.~\ref{scheme_full}.}
\label{scheme_simple}
\end{figure}

Third, the exonuclease site can only excise the terminal nucleotide. What happens after the cleavage is not clear yet\cite{lamichhane2013dynamics}. Here we propose two possible pathways, which are denoted as Model I and Model II in FIG.~\ref{scheme_full}. In Model I, DNAP undergoes a backward translocation and the primer terminus can either be excised processively, or be transferred back to the polymerase site (at the pre-translocation state).
In Model II, the primer terminus is directly transferred back to the polymerase site (at the post-translocation state).
FIG. \ref{scheme_full} can be further simplified as FIG.~\ref{scheme_simple}, considering that the addition of dNTP in the polymerase site is almost irreversible (i.e., the product PPi of the polymerization reaction is often released irreversibly under physiological conditions).

One can also reasonably assume that the translocation of DNAP in `polymerase type' complex is in a rapid equilibrium. In biochemical experimental studies such as steady-state kinetic assays\cite{donlin1991kinetic,tsai2006new}, the translocation cannot be observed (for comparison, the subsequent dNTP binding can be clearly observed). In other words, the two substates can not be identified individually, indicating there exists a rapid equilibrium between them.  Thus one does not need to distinguish between the pre-translocation and the post-translocation states. Under such an approximation, Model II can be reduced to the Galas-Branscomb model as shown in FIG.~\ref{scheme_compare}(b1),
while Model I is reduced to FIG.~\ref{scheme_compare}(c1).

Although Model II were widely accepted, there is no direct experimental evidence to exclude Model I. Moreover, it has been found that the ssDNA binding to the exonuclease site can be processively excised\cite{donlin1991kinetic}, indicating that more than one nucleotide bind to the exonuclease site (e.g., three nucleotides bind to the exonuclease site for Polymerase I KF\cite{beese1993structure}) and removing the terminal nucleotide may trigger backward translocation of DNAP for the subsequent excisions.
So we will discuss both models in this paper, but put a focus on Model I due to the following technique consideration.
Kinetic proofreading models like FIG.~\ref{scheme_compare}(a1) or (a2) are irreversible reactions,
so the corresponding kinetic equations are always closed (i.e., of finite number) and can be rigorously calculated.
The Galas-Branscomb models like FIG.~\ref{scheme_compare}(b1) or (b2), however, are seemingly reversible, and the corresponding kinetic equations are always unclosed and hierarchically coupled, which is hard to solve.
Fortunately, a general rigorous treatment for such problems has been established recently by us\cite{shu2015general}, and this method can be directly applied to Model II (some important results are given in Appendix \ref{app:m2}).
For Model I like FIG.~\ref{scheme_compare}(c1) or (c2), however, the above methods are inapplicable and new method should be developed, which will be a focus of later sections.

\begin{figure}
\includegraphics[width=.5\textwidth]{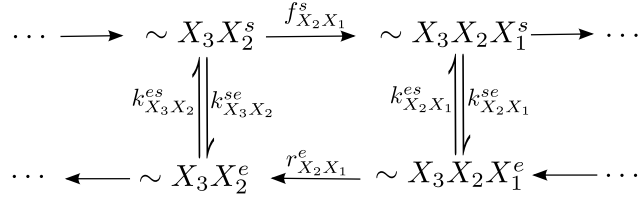}
\caption{The minimal scheme of the first-order proofreading model.}
\label{model1_order1}
\end{figure}

This paper is organized as follows. Section \ref{sec:theory} introduces the basic theory of the steady-state kinetics of Model I (the minimal scheme FIG.~\ref{scheme_compare}(c2)) including higher-order neighbor effects.
In Section \ref{sec:fidelity}, we discuss the replication fidelity problem of DNAP with either Model I or Model II (FIG.~\ref{scheme_compare}(c2),(b2) respectively).
While it's hard to analytically calculate the fidelity in terms of the kinetic parameters from the basic kinetic theory, we introduce an alternative method (infinite-state Markov chain) for the calculation and show numerically its equivalence to our basic theory.
With this method, analytical expressions for fidelity are obtained under the so-called biologically-relevant conditions.
We further show that Model I and II give exactly the same expressions which offer an intuitive understanding of the higher-order neighbor effects on the fidelity.
In Section \ref{application}, we will apply these results to discuss the fidelity problem of some real DNAPs.

\section{Basic kinetic theory of proofreading model I} \label{sec:theory}

It has been shown that the terminal mismatch and even the penultimate mismatch at the primer terminus will greatly reduce the addition rate of the next dNTP, compared with the case that a match is at the same position \cite{wong1991induced,johnson1993conformational,johnson2001fidelity}.
This means that some rate constants in FIG.~\ref{scheme_full} depends on the states ($A$ or $B$) of the few consecutive base pairs at the terminal region, i.e., there does exist higher-order neighbor effects (referred to as terminal effects in this paper) in DNA replication.
Thus the zero-order terminal model shown in FIG.~\ref{scheme_simple} is not appropriate and higher-order models like FIG.~\ref{model1_order1} or FIG.~\ref{model1_order2} are required.
Below we demonstrate how to rigorously treat the steady-state kinetics of such models.
To proceed, we note first that each step in the reaction scheme may have terminal effect but of different order. For instance, the addition rate may be of first order while the transfer rate may be of zero order, which is a special case of the general first-order scheme FIG.~\ref{model1_order1} (by putting $k^{se}_{AX_1} =k^{se}_{BX_1}$).
Similarly, reaction schemes with kinetic parameters up to $s$th order can be included in the general $s$th-order scheme.

\subsection{First-order proofreading model}
In this subsection, we will discuss the general first-order proofreading model FIG.~\ref{model1_order1} to demonstrate the basic ideas of our approach. Following the same logic of Ref.~\onlinecite{shu2015general}, we use $P_{X_n\cdots X_1}^s$ to denote the occurrence probability of the terminal sequence $X_n\cdots X_1$ in the synthetic (polymerase) site, $P_{X_n\cdots X_1}^e$ to denote the occurrence probability of $X_n\cdots X_1$ in the exonuclease site, $X_i = A, B$.
$N_{X_n\cdots X_2X_1}$ is defined as the total number of sequence $X_n\cdots X_2X_1$ appearing in the primer chain.

The overall incorporation rate of sequence $X_n\cdots X_2X_1(n\ge 2)$ is defined as,
\begin{equation}
\dot{N}_{X_n\cdots X_2X_1}\equiv J_{X_n\cdots X_2X_1}=J_{X_n\cdots X_2X_1}^s+J_{X_n\cdots X_2X_1}^e,
\end{equation}
where $J_{X_n\cdots X_2X_1}^s=f_{X_2X_1}^s P_{X_n\cdots X_2}^s$, $J_{X_n\cdots X_2X_1}^e=-r_{X_2X_1}^e P_{X_n\cdots X_2X_1}^e$.

The kinetic equations of $P_{X_n\cdots X_2 X_1}^m(n\ge 1,m=s,e)$ can be written as,
\begin{subequations}
\begin{eqnarray}
\dot{P}_{X_n\cdots X_2 X_1}^s=J_{X_n\cdots X_2 X_1}^s-\tilde{J}_{X_n\cdots X_2 X_1\ast}^s-J_{X_n\cdots X_2 X_1}^{se},\\
\dot{P}_{X_n\cdots X_2 X_1}^e=J_{X_n\cdots X_2 X_1}^e-\tilde{J}_{X_n\cdots X_2 X_1\ast}^e+J_{X_n\cdots X_2 X_1}^{se},
\end{eqnarray}
\end{subequations}
where, $\tilde{J}_{X_n\cdots X_1\ast}^s= J_{X_n\cdots X_1A}^s+J_{X_n\cdots X_1B}^s$, $\tilde{J}_{X_n\cdots X_1\ast}^e= J_{X_n\cdots X_1A}^e+J_{X_n\cdots X_1B}^e$, $J_{X_n\cdots X_2 X_1}^{se}=k_{X_2X_1}^{se} P_{X_n\cdots X_2X_1}^s-k_{X_2X_1}^{es} P_{X_n\cdots X_2X_1}^e$. We also have $P_{X_i\cdots X_1}^s=P_{AX_i\cdots X_1}^s + P_{B X_i\cdots X_1}^s$, $J_{X_i\cdots X_1}^s=J_{AX_i\cdots X_1}^s + J_{B X_i\cdots X_1}^s$ ($i \geq 1$) and so on.

For example,
\begin{subequations}
\begin{eqnarray}
\dot{P}_{AB}^s &=&f_{AB}^s(P_{AA}^s+P_{BA}^s)-(f_{BA}^s+f_{BB}^s)P_{AB}^s-k_{AB}^{se}P_{AB}^s+k_{AB}^{es}P_{AB}^e \nonumber \\
&=&J_{AB}^s-(J_{ABA}^s+J_{ABB}^s)-J_{AB}^{se},\\
\dot{P}_{AB}^e & = & -r_{AB}^eP_{AB}^e+r_{BA}^eP_{ABA}^e+r_{BB}^eP_{ABB}^e+k_{AB}^{se}P_{AB}^s-k_{AB}^{es}P_{AB}^e \nonumber \\
&=&J_{AB}^e-(J_{ABA}^e+J_{ABB}^e)+J_{AB}^{se}.
\end{eqnarray}
\end{subequations}

The steady state is defined as $\dot{P}_{X_n\cdots X_2 X_1}^s=0$ and $\dot{P}_{X_n\cdots X_2 X_1}^e=0$ for any $n\ge 1$. To rigorously solve these coupled equations, we extend the logic of Ref.~\onlinecite{shu2015general} and propose the following factorization conjecture:
\begin{eqnarray}
P_{X_n\cdots X_2X_1}^m=\prod_{i=3}^{n}{P_{X_iX_{i-1}}^s} \biggl[\prod_{i=3}^{n}P_{X_{i-1}}^s\biggr]^{-1} P_{X_2X_1}^m, \quad n\ge 3,  \quad m=s,e.
\end{eqnarray}

\begin{figure}
\includegraphics[width=.8\textwidth]{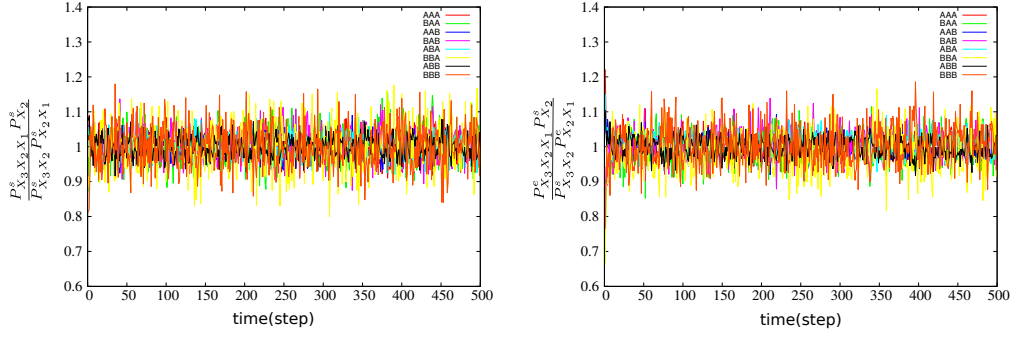}
\caption{Simulation verification of the factorization conjecture of the first-order proofreading model, with illustrative rate parameters(in unit $s^{-1}$) $f_{AA}^s=8$, $f_{AB}^s=6$, $f_{BA}^s=4$, $f_{BB}^s=2$, $r_{AA}^e=1$, $r_{AB}^e=1$, $r_{BA}^e=2$, $r_{BB}^e=1$, $k_{AA}^{se}=1$, $k_{AB}^{se}=6$, $k_{BA}^{se}=1$, $k_{BB}^{se}=6$, $k_{AA}^{es}=1$, $k_{AB}^{es}=3$, $k_{BA}^{es}=1$, $k_{BB}^{es}=4$. Averaged over 10,000 samples.}
\label{order1_simu}
\end{figure}

For example, $P_{X_3X_2X_1}^s=P_{X_3X_2}^sP_{X_2X_1}^s/P_{X_2}^s$, $P_{X_3X_2X_1}^e=P_{X_3X_2}^sP_{X_2X_1}^e/P_{X_2}^s$ (correspondingly, one also has $J_{X_3X_2X_1}^s=P_{X_3X_2}^sJ_{X_2X_1}^s/P_{X_2}^s$ and $J_{X_3X_2X_1}^e=P_{X_3X_2}^sJ_{X_2X_1}^e/P_{X_2}^s$ ).
The validity of these factorization conjectures can be numerically tested by Monte Carlo simulation (using the Gillespie algorithm\cite{gillespie1977exact}), as shown in FIG.~\ref{order1_simu}.

By this factorization conjecture, the original unclosed equations can be reduced to the following closed equations of the eight basic variables $P_{X_2X_1}^{m}(m=s,e)$:
\begin{eqnarray}\label{1steqs}
\nonumber
J_{BA}^e-J_{AB}^e=J_B^{se}, \ \ \ \ \ \  J_{BA}^s-J_{AB}^s=J_A^{se}, \\ \nonumber
\frac{J_{AA}^s-J_{AA}^{se}}{J_{BA}^s-J_{BA}^{se}}=\frac{P_{AA}^s}{P_{BA}^s}, \ \ \ \frac{J_{AB}^s-J_{AB}^{se}}{J_{BB}^s-J_{BB}^{se}}=\frac{P_{AB}^s}{P_{BB}^s}, \\
\frac{J_{AA}^e+J_{AA}^{se}}{J_{BA}^e+J_{BA}^{se}}=\frac{P_{AA}^s}{P_{BA}^s}, \ \ \
 \frac{J_{AB}^e+J_{AB}^{se}}{J_{BB}^e+J_{BB}^{se}}=\frac{P_{AB}^s}{P_{BB}^s},\\ \nonumber
J_A^{se}+J_B^{se}=0, \ \ \ \  \sum\limits_{X,Y=A,B} (P_{XY}^s+P_{XY}^e)=1. \nonumber
\end{eqnarray}

\subsection{Second-order proofreading model}
\begin{figure}
\includegraphics[width=.6\textwidth]{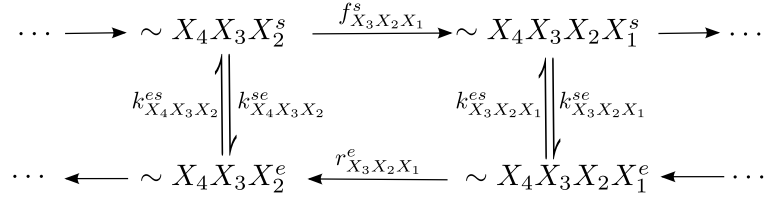}
\caption{The second-order proofreading reaction scheme.}
\label{model1_order2}
\end{figure}

Second-order terminal effects have been observed for some DNAPs where the penultimate mismatch at the terminus can affect the next nucleotide incorporation \cite{johnson2001exonuclease,johnson2001fidelity}. In this section, we extend the method of the preceding subsection to the second-order model shown in FIG.~\ref{model1_order2}.

Similar to the first-order model, we have,
\begin{equation}
\dot{N}_{X_n\cdots X_2X_1}\equiv J_{X_n\cdots X_2X_1}=J_{X_n\cdots X_2X_1}^s+J_{X_n\cdots X_2X_1}^e \quad (n\ge 3),
\end{equation}
where $J_{X_n\cdots X_3X_2X_1}^s=f_{X_3X_2X_1}^s P_{X_n\cdots X_3X_2}^s$, $J_{X_n\cdots X_3X_2X_1}^e=-r_{X_3X_2X_1}^e P_{X_n\cdots X_3X_2X_1}^e$.

The kinetic equations for $P_{X_n\cdots X_3X_2 X_1}^m(n\ge 1,m=s,e)$ can be written as,
\begin{subequations}
\begin{eqnarray}
\dot{P}_{X_n\cdots X_3X_2 X_1}^s=J_{X_n\cdots X_3X_2 X_1}^s-\tilde{J}_{X_n\cdots X_3X_2 X_1\ast}^s-J_{X_n\cdots X_3X_2 X_1}^{se}, \\
\dot{P}_{X_n\cdots X_3X_2 X_1}^e=J_{X_n\cdots X_3X_2 X_1}^e-\tilde{J}_{X_n\cdots X_3X_2 X_1\ast}^e+J_{X_n\cdots X_3X_2 X_1}^{se},
\end{eqnarray}
\end{subequations}
where $J_{X_n\cdots X_3X_2 X_1}^{se}= k_{X_3X_2X_1}^{se} P_{X_n\cdots X_3X_2X_1}^s-k_{X_3X_2X_1}^{es} P_{X_n\cdots X_3X_2X_1}^e$.

Under steady-state conditions $\dot{P}_{X_n\cdots X_3X_2 X_1}^s=\dot{P}_{X_n\cdots X_3X_2 X_1}^e=0$, we proposed the following factorization conjecture:
\begin{eqnarray}
P_{X_n\cdots X_3X_2X_1}^m=\prod_{i=4}^{n}{P_{X_iX_{i-1}X_{i-2}}^s} \biggl[\prod_{i=4}^{n}P_{X_{i-1}X_{i-2}}^s\biggr]^{-1} P_{X_3X_2X_1}^m, \quad n\ge 4, \quad m=s,e,
\end{eqnarray}
which can be tested by Monte Carlo simulations (results not shown here).

Therefore, we obtain the following closed equations for the second-order proofreading model:
\begin{eqnarray} \label{2steqs}
\nonumber
J_{X\bar{X}}^s-\tilde{J}_{X\bar{X}\ast}^s=J_{X\bar{X}}^{se},  &  \ \  J_{\bar{X}XX}^s-J_{XX\bar{X}}^s=J_{XX}^{se},\\ \nonumber
\frac{J_{AXY}^s-J_{AXY}^{se}}{J_{BXY}^s-J_{BXY}^{se}}=\frac{P_{AXY}^s}{P_{BXY}^s}, & J_{XX\bar{X}}=J_{\bar{X}XX}, \\
\frac{J_{AXY}^e+J_{AXY}^{se}}{J_{BXY}^e+J_{BXY}^{se}}=\frac{P_{AXY}^s}{P_{BXY}^s}, & J_{AB}=J_{BA}, \\ \nonumber
\sum(P_{XYZ}^s+P_{XYZ}^e)=1,  &   X,Y,Z=A,B,    \nonumber
\end{eqnarray}
where $\bar{X}$ differs from $X$.

Some experiments\cite{johnson1993conformational,johnson2004structures} show that up to 4 base pairs at the primer terminus may have apparent effects on the incorporation rates of the next nucleotide. For such cases, one should generalize the above method to include these higher-order terminal effects. The generalization to $s$th-order model is straightforward and details are not given here.

\section{The fidelity problem of DNA replication by DNAP} \label{sec:fidelity}
In this section, we discuss the fidelity problem of DNAP.  In principle, one can define the fidelity naturally as the ratio of matches over mismatches incorporated into the primer.
However, it's difficult to directly measure this fidelity in experiments and some indirect methods were developed.
One of the common used methods is the forward mutation assay\cite{tindall1988fidelity,Lundberg1991High,Cline1996PCR,kokoska2002low} which scores the replication errors indirectly by counting the phenotype change rate of the bacterial hosts transfected by reporter gene DNA.
Other frequently used methods are steady-state\cite{boosalis1987dna,goodman1993biochemical,bebenek1990fidelity} or pre-steady state\cite{kuchta1987kinetic,kuchta1988kinetic,patel1991pre,fiala2004pre} kinetic assays
which investigate the kinetics of DNA replication and calculate the replication fidelity indirectly based on the theoretical models.
The basic ideas of these two approaches differ, but the obtained fidelity are often of similar order of magnitudes. For example, the average fidelity of Sulfolobus solfataricus P2 DNA polymerase IV (Dpo4) is about $1.3 \times 10^2$ to $3.3\times 10^3$ using steady-state kinetic method\cite{boudsocq2001sulfolobus}, which agrees with  $1.5 \times 10^2$ given by forward mutation assay \cite{kokoska2002low}. In general, for most proofreading-proficient DNAPs, the fidelity \emph{in vitro} is about $10^6-10^7$ with a contribution by exonuclease proofreading of $10^1-10^2$.\cite{kunkel1992dna,kunkel2000dna}

In this paper, we only discuss the kinetic-based fidelity, since it can be rigorously defined and calculated within the framework of our basic theory. Here we define the fidelity as $\phi=N_A/N_B$.  $N_A$ is the total number of incorporated matches in the primer, $N_B$ is the total number of mismatches.
Once the steady-state kinetic equations such as Eqs.~(\ref{1steqs}) or Eqs.~(\ref{2steqs}) are solved numerically or analytically, the total flux $J_A (=J^s_A + J^e_A)$, $J_B (=J^s_B + J^e_B)$ can be calculated.
Since $\dot{N}_A=J_A$, $\dot{N}_B=J_B$, and $d(N_A/N_B)/dt=0$ (in steady state),  we can calculate the replication fidelity exactly by $\phi=N_A/N_B=J_A/J_B$.
In particular, the analytical solutions to the kinetic equations are quite useful for further experimental and theoretical studies.
However, it's often impossible to solve the kinetic equations analytically.
To circumvent this problem, we introduce below an alternative method, the infinite-state Markov chain mehod \cite{cady2009open}, to calculate $\phi$.
This method has already been used for higher-order copolymerization by us (see the supplementary of Ref.~\onlinecite{shu2015general}) and can be readily extended to the exonuclease proofreading schemes.

\subsection{The infinite-state Markov chain method for exonuclease proofreading}
\begin{figure}
\includegraphics[width=.7\textwidth]{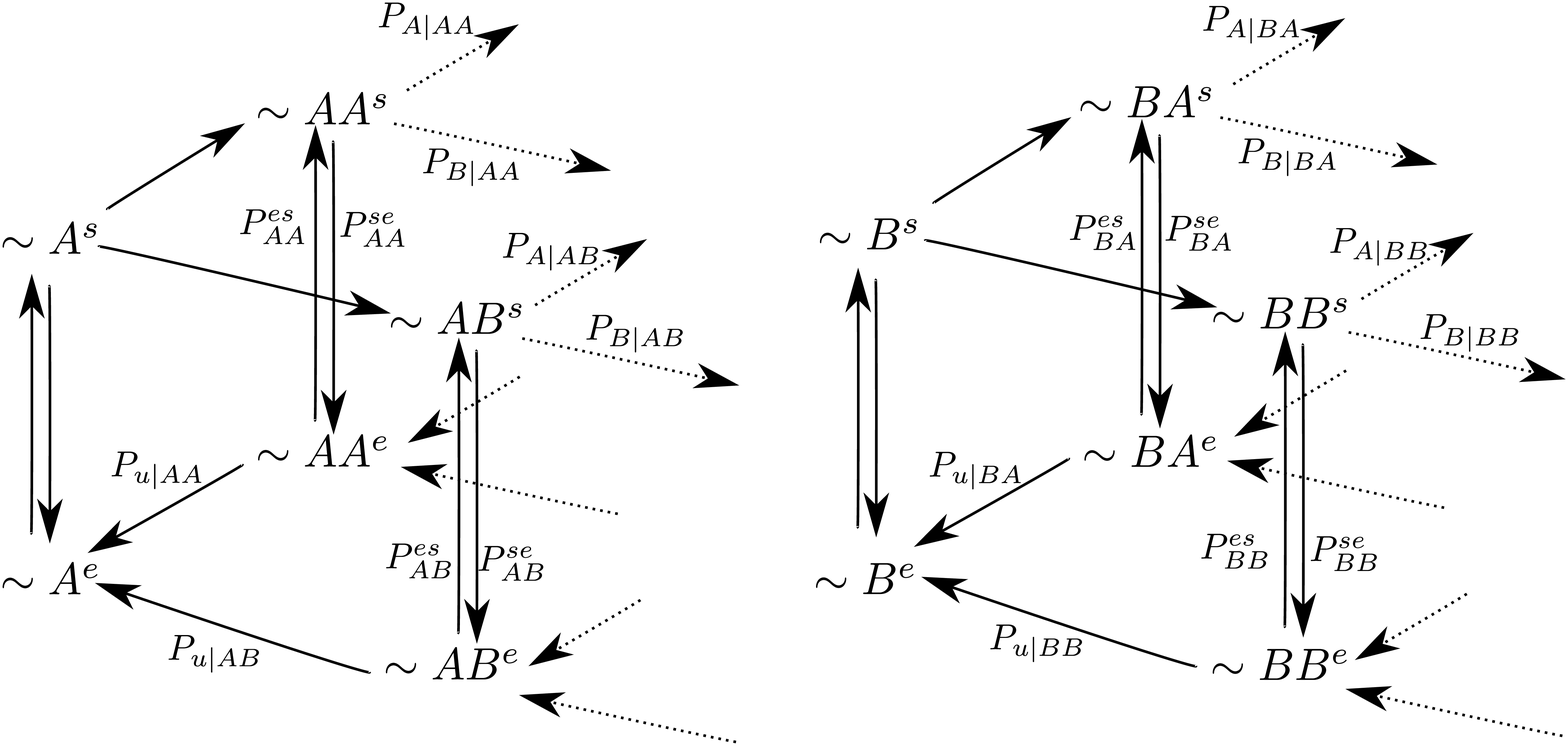}
\caption{Branching model for the first-order polymerization and excision.}
\label{branching}
\end{figure}

To calculate the fidelity, we begin with the first-order proofreading scheme which can be rewritten as a branching model shown in FIG.~\ref{branching}.

The steady-state growth of primer can be completely characterized by four groups of transition probabilities:
\begin{eqnarray*} \label{eq:fid}
&&P_{X|X_2X_1}\equiv f_{X_1X}^s/(f_{X_1X}^s+f_{X_1\bar{X}}^s+k_{X_2X_1}^{se}),  \quad
P_{X_2X_1}^{se}\equiv k_{X_2X_1}^{se}/(f_{X_1X}^s+f_{X_1\bar{X}}^s+k_{X_2X_1}^{se}),  \\
&&P_{X_2X_1}^{es}\equiv k_{X_2X_1}^{es}/(r_{X_2X_1}^e+k_{X_2X_1}^{es}),  \quad
P_{u|X_2X_1}\equiv 1-P_{X_2X_1}^{es} = r_{X_2X_1}^e/(r_{X_2X_1}^e+k_{X_2X_1}^{es}).
\end{eqnarray*}

We also employ the idea of `cycle completion' \cite{cady2009open}, since any incorporated nucleotide (either $A$ or $B$) has a chance to be excised, only those not being excised account for the final composition of the primer. Thus the fidelity for the first-order terminal model can be defined as,
\begin{eqnarray} \label{eq:fid1}
\phi\equiv \frac{Q_{AA} + Q_{BA} }{Q_{AB} + Q_{BB}},
\end{eqnarray}
where $Q_{X_2X_1}$ is the probability that $X_1$ is added to the terminal $X_2$ and never being excised, satisfying $Q_{AA}+Q_{AB}+Q_{BA}+Q_{BB}=1$. $Q_{X_2X_1}$ can be explicitly expressed as $Q_{X_2X_1}\equiv \hat{P}_{X_2X_1} P_{nuX_2X_1}$, where $\hat{P}_{X_2X_1}$ is the probability that adding $X_1$ to the terminal $X_2$, $P_{nuX_2X_1}$ is the probability of the terminal $X_2X_1$ never being excised.
The absolute values of $\hat{P}_{X_2X_1}$ are not known \textit{a prior}, but the following equalities obviously hold:
\begin{eqnarray} \label{eq:split}
\frac{\hat{P}_{AA}}{\hat{P}_{AB}}=\frac{P_{A|AA}}{P_{B|AA}}=\frac{P_{A|BA}}{P_{B|BA}}=\frac{f_{AA}^s}{f_{AB}^s}, \ \ \frac{\hat{P}_{BA}}{\hat{P}_{BB}}=\frac{P_{A|AB}}{P_{B|AB}}=\frac{P_{A|BB}}{P_{B|BB}}=\frac{f_{BA}^s}{f_{BB}^s}.
\end{eqnarray}

Considering the fact that the number of $AB$ should equal to the number of $BA$ in the copolymer chain, we have the following intrinsic constraint:
\begin{eqnarray}\label{eq:r2}
Q_{AB}(=\hat{P}_{AB} P_{nuAB})=Q_{BA}(=\hat{P}_{BA} P_{nuBA}).
\end{eqnarray}

To calculate $P_{nuX_2X_1}$, we define $P_{euX_2X_1} \equiv 1- P_{nuX_2X_1}$ as the probability of the terminal $X_2X_1$ ever being excised. $P_{euX_2X_1}$ satisfy the following iterative equations (details can be found in Appendix \ref{app:iter}):
\begin{equation} \label{eq:iter}
P_{euX_2X_1}  = \frac{\hat{P}_{u|X_2X_1}}{P_{X_2X_1}^{se}P_{X_2X_1}^{es}}\left(\frac{1}{1-(\hat{P}_{A|X_2X_1}P_{euX_1A}+\hat{P}_{B|X_2X_1}P_{euX_1B})P_{X_2X_1}^{es}}-\frac{1}{T_{X_2X_1}}\right).
\end{equation}

Here, $T_{X_2X_1}=1/(1-P_{X_2X_1}^{se}P_{X_2X_1}^{es})$, $\hat{P}_{u|X_2X_1}=P_{u|X_2X_1}P_{X_2X_1}^{se}T_{X_2X_1}=\hat{r}_{X_2X_1}^s/(f_{X_1A}^s+f_{X_1B}^s+\hat{r}_{X_2X_1}^s)$,
$\hat{P}_{X|X_2X_1}=P_{X|X_2X_1}T_{X_2X_1}=f_{X_1X}^s/(f_{X_1X}^s+f_{X_1\bar{X}}^s+\hat{r}_{X_2X_1}^s)$,
$\hat{r}_{X_2X_1}^s=k_{X_2X_1}^{se}r_{X_2X_1}^e/(r_{X_2X_1}^e+k_{X_2X_1}^{es})$.

Once $P_{euX_2X_1}$ are solved,  $\hat{P}_{X_2X_1}$ can then be calculated by combing Eq.~(\ref{eq:split}) and Eq.~(\ref{eq:r2}), and the fidelity $\phi$ can be obtained by Eq.~(\ref{eq:fid1}).
Numerical calculations show that $\phi$ obtained in this approach is identical to that given by the steady-state kinetic equations Eqs.~(\ref{1steqs}), which provides a verification of our kinetic approach.
The same logic can be extended to any higher-order models.

\subsection{Approximation of $\phi$ under bio-relevant conditions} \label{sec:approx}
In TABLE~\ref{table1}, we list experimental values of some kinetic parameters for some real DNAPs.
There exists huge difference in the order of magnitudes of the parameters of the same DNAP. For example, addition of matched nucleotide at the polymerase site is very fast, and always much faster than mismatch addition.
This enables us to suggest reasonable approximations (so-called bio-relevant conditions in this paper) to simplify the above calculation and obtain explicit mathematical expressions of $\phi$ in terms of some key parameters.
For any higher-order models (say, $h^{th}$-order model), we propose

\begingroup
\squeezetable
\begin{table}
\caption{Experimental values of kinetic parameters of some real DNAPs}
\begin{ruledtabular}
\begin{tabular}{|c|c|c|c|c|c|c|}
\multicolumn{2}{|c|}{rate parameter} & T7\cite{wong1991induced,donlin1991kinetic} ($s^{-1}$) & pol $\gamma$\cite{johnson2001fidelity,johnson2001exonuclease} ($s^{-1}$) & Pol III\cite{bloom1997fidelity,miller1996kinetic} ($s^{-1}$) & T4\cite{capson1992kinetic} ($s^{-1}$) & phi29 DNAP\cite{esteban1993fidelity,esteban19943,lieberman2014kinetic} ($s^{-1}$)\\
\hline
 & $f_{AA}^s$ & 250\footnotemark[1] & 3900-5700\footnotemark[3] & 370\footnotemark[5] & 600 & 680\footnotemark[7]\\
  & $f_{AB}^s$ & 0.002\footnotemark[1] & 0.023-1.6\footnotemark[3] & (0.16-2.1)$\times 10^{-3}$\footnotemark[5] & $\ast$ & $10^{-4}$-$10^{-6}$\footnotemark[7] \\
first & $k_{AA}^{se}$ & 0.2\footnotemark[2] & $>\sim$0.05\footnotemark[4] & 0.015\footnotemark[6] & 1  & 11.54\footnotemark[9]\\
 & $k_{AB}^{se}$ & 2.3\footnotemark[2] & $>\sim$0.4\footnotemark[4] & 0.038\footnotemark[6] & 5 & -- \\
 order & $k_{AA}^{es}$ & 714\footnotemark[2] & $>\sim$39\footnotemark[4] & -- & 20 & 10.48\footnotemark[9] \\
 & $k_{AB}^{es}$ & 714\footnotemark[2] & -- & -- & -- & --\\
 & $r^e$ & 896\footnotemark[2] & $>\sim$39\footnotemark[4] & 280\footnotemark[6] & 100 & 500\footnotemark[8] \\
\hline
  & $k_{ABA}^{se}$ & -- & $>\sim$3\footnotemark[4] & -- & -- & -- \\
 second & $k_{ABA}^{es}$ & -- & -- & -- & -- & --\\
 order& $f_{ABA}^s$ & 0.012\footnotemark[1] & 0.1\footnotemark[3] & -- & --  & $10^{-3}-10^{-4}$\footnotemark[7]\\
 & $f_{BAA}^s$ & -- & 2.7\footnotemark[3] & -- & -- & -- \\
\end{tabular}
\end{ruledtabular}
\footnotetext[0]{Polymerization parameters are all scaled to the standard dNTP concentration $100\mu M$. \hspace{16ex} -- means the data were not found. \ \ \ $\ast$ means the data is too small to measure.}
\footnotetext[1] {from Table II of Ref.~\onlinecite{wong1991induced}.} \footnotetext[2] {from Ref.~\onlinecite{donlin1991kinetic}.} \footnotetext[3] {from Ref.~\onlinecite{johnson2001fidelity}, dNTP concentration is set as $100\mu M$ for holoenzyme. The listed values differ for different base pairs (matched or mismatched). This type of sequence effect is beyond the scope of this paper and not discussed here. } \footnotetext[4] {estimated from the combined kinetic parameters from Ref.~\onlinecite{johnson2001exonuclease}.}
\footnotetext[5] {values for the holoenzyme, from Ref.~\onlinecite{bloom1997fidelity}.}
\footnotetext[6] {estimated by pre-steady state measurements of purified $\epsilon$ subunit, from Ref.~\onlinecite{miller1996kinetic}.}
\footnotetext[7] {from Table I and Table II of Ref.~\onlinecite{esteban1993fidelity} in Mg$^{2+}$-activated polymerization.} \footnotetext[8] {from Ref.~\onlinecite{esteban19943}.}
\footnotetext[9] {from Ref.~\onlinecite{lieberman2014kinetic}.}
\label{table1}
\end{table}
\endgroup

(a) $f_{\underbrace{\scriptscriptstyle{AAA}\cdots}_{h+1}}^s\gg f_{\underbrace{\scriptscriptstyle{AAA}\cdots}_{h}B}^s$, which leads to $P_{A|X\underbrace{\scriptscriptstyle{AAA}\cdots}_{h}}\gg P_{B|X\underbrace{\scriptscriptstyle{AAA}\cdots}_{h}}$.

This means that the overall nucleotide incorporation is dominated by the addition of $A$ and the occurrence probability of $B$ in the primer is negligible. This highly efficient discrimination between $A$ and $B$ is executed by the polymerase site.

(b)$f_{\underbrace{\scriptscriptstyle{AAA}\cdots}_{h+1}}^s\gg k_{X\underbrace{\scriptscriptstyle{AAA}\cdots}_{h}}^{se}(>\hat{r}_{X\underbrace{\scriptscriptstyle{AAA}\cdots}_{h}}^s)$, which leads to $P_{A|X\underbrace{\scriptscriptstyle{AAA}\cdots}_{h}}\gg P^{se}_{X\underbrace{\scriptscriptstyle{AAA}\cdots}_{h}} (>\hat{P}_{u|X\underbrace{\scriptscriptstyle{AAA}\cdots}_{h}})$.

This can be achieved at appropriate concentration of dNTP (notice that $f_{\underbrace{\scriptscriptstyle{AAA}\cdots}_{h+1}}^s$ is proportional to dNTP concentration). It means that the matched terminus can be rapidly extended by the next match, instead of being transferred to the exonuclease site and excised. This ensures that the primer growth is dominated by match extension in the polymerase site and the introduction of exonuclease proofreading pathway nearly does not change the overall growth velocity.

(c)$\hat{r}_{\underbrace{\scriptscriptstyle{AAA}\cdots}_{h-i+1}B\underbrace{\scriptscriptstyle{AAA}\cdots}_{i-1}}^s \agt f_{\underbrace{\scriptscriptstyle{AAA}\cdots}_{h-i}B\underbrace{\scriptscriptstyle{AAA}\cdots}_{i}}^s$ for $0< i \leq m$
(which leads to $\hat{P}_{u|\underbrace{\scriptscriptstyle{AAA}\cdots}_{h-i+1}B\underbrace{\scriptscriptstyle{AAA}\cdots}_{i-1}}(\equiv R_i) \agt \hat{P}_{A|\underbrace{\scriptscriptstyle{AAA}\cdots}_{h-i+1}B\underbrace{\scriptscriptstyle{AAA}\cdots}_{i-1}}(\equiv F_i)$),
and $\hat{r}_{\underbrace{\scriptscriptstyle{AAA}\cdots}_{h-i+1}B\underbrace{\scriptscriptstyle{AAA}\cdots}_{i-1}}^s \ll f_{\underbrace{\scriptscriptstyle{AAA}\cdots}_{h-i}B\underbrace{\scriptscriptstyle{AAA}\cdots}_{i}}^s$ for $m < i \leq h$ (i.e., $R_i \ll F_i$). $m$ is an arbitrary integer in the range [0,h].

This means that the primer terminus containing a mismatch is more readily transferred and excised rather than extended by the addition of the next matched nucleotide. This makes a significant contribution to the proofreading efficiency. On the other hand, as the mismatch is buried deeper (i.e., $i$ gets larger),
the transfer-and-excision rate $\hat{r}_{\underbrace{\scriptscriptstyle{AAA}\cdots}_{h-i+1}B\underbrace{\scriptscriptstyle{AAA}\cdots}_{i-1}}^s$
decreases and the addition rate
$f_{\underbrace{\scriptscriptstyle{AAA}\cdots}_{h-i}B\underbrace{\scriptscriptstyle{AAA}\cdots}_{i}}^s$
increases and far exceeds the transfer-and-excision rate when $i>m$.
Hence, only those kinetic parameters of $0 < i \le m$ contribute significantly to the proofreading efficiency. More details about this condition can be found in Appendix \ref{app:approx}.

(d) $f_{X_{h}\cdots X_1B}^s \approx 0$ (where $X_{h}X_{s-1}\cdots X_1\neq \underbrace{{AAA}\cdots}_{h}$), which leads to $P_{B|X_{h+1}X_{h}\cdots X_1}=0$.

This means that the chance of adding one more mismatch within the length of $h$ is negligible.

With these bio-relevant conditions, a very simple and intuitive expression of the replication fidelity can be obtained:
\begin{eqnarray} \label{eq:fid_tot}
 \phi= \phi_s \phi_e, \quad \phi_s=f_{\underbrace{\scriptscriptstyle{AAA\cdots}}_{h+1}}^s/f_{\underbrace{\scriptscriptstyle{AAA\cdots}}_{h}B}^s, \nonumber \\
\phi_e= \left( 1+\frac{R_1}{F_1}\right) \left( 1+\frac{R_2}{F_2}\right) \left(\cdots\right)\left( 1+\frac{R_h}{F_h} \right).
\end{eqnarray}

$\phi_s, \phi_e$ denotes the contribution of the polymerase pathway and the proofreading pathway to the overall fidelity, respectively (details can be found in Appendix \ref{app:approx}).

Particularly, for the first-order model, we have,
\begin{eqnarray} \label{exp:first}
\phi_s=\frac{f_{AA}^s}{f_{AB}^s},	\quad \phi_e=1+\frac{\hat{r}_{AB}^s}{f_{BA}^s}.
\end{eqnarray}

For the second-order model, we have,
 \begin{eqnarray} \label{exp:second}
\phi_s=\frac{f_{AAA}^s}{f_{AAB}^s},  \quad \phi_e=(1+\frac{\hat{r}_{AAB}^s}{f_{ABA}^s})(1+\frac{\hat{r}_{ABA}^s}{f_{BAA}^s}).
\end{eqnarray}
Here  $\hat{r}_{X_3X_2X_1}^s=k_{X_3X_2X_1}^{se}r_{X_3X_2X_1}^e/(r_{X_3X_2X_1}^e+k_{X_3X_2X_1}^{es})$, similarly defined as  $\hat{r}_{X_2X_1}^s$ for the first-order model.
If all the parameters are taken as first order, the term $\hat{r}_{ABA}^s/f_{BAA}^s$ becomes negligible (according to the condition (b) $f_{AA}^s \gg \hat{r}_{BA}^{s}$), and Eq.~(\ref{exp:second}) is indeed reduced to Eq.~(\ref{exp:first}).

For Model II, following similar procedure, one can derive the same expression of $\phi$ as Eq.~(\ref{eq:fid_tot}) under the same conditions (details see Appendix \ref{app:m2}).
Furthermore, by numerically solving the steady-state kinetic equations (e.g., Eqs.~(\ref{1steqs})), one can also show that Model I and II give almost the same overall reaction velocity ($J_{tot}=J_A + J_B$) under the bio-relevant conditions (data not shown).
This is conceivable, since the overall velocity is dominated by the addition of $A$ ($f_{\underbrace{\scriptscriptstyle{AAA}\cdots}_{h+1}}^s$ is far larger than any other kinetic parameters)
and introduction of proofreading pathway only slightly changes the overall velocity.
Therefore, the two models behave almost the same in steady state under the bio-relevant conditions (they do differ under other conditions, which beyond the topic of the present paper).
This means that the details how the excised terminus returns to the polymerase site may be unimportant for real DNAPs to obtain high proofreading efficiency while maintain high polymerization velocity.

\section{Case studies} \label{application}
In the above expressions of $\phi$, only a few key parameters appear, which enables us to evaluate the fidelity of some real DNAPs even if other unimportant kinetic parameters are unknown or not precisely measured. Here we give two case studies.

\subsection{First-order proofreading}

Employing the pre-steady-state kinetic analysis method, K.A.Johnson et al. analyzed the polymerization process and the excision process of T7 DNA polymerase\cite{patel1991pre,wong1991induced,donlin1991kinetic}. The kinetic parameters they obtained are listed in TABLE~\ref{table1}, and can be understood as first order parameters. Since they satisfy the  bio-relevant conditions, Eq.~(\ref{exp:first}) can be applied here.

For $\phi_s$, K.A.Johnson et al. used an expression exactly the same as ours ($=f_{AA}^s/f_{AB}^s \simeq 10^5$). However, for $\phi_e$, they calculated as	
 \begin{equation} \label{johnson1}
 \phi_e=1+\frac{k_{AB}^{se}}{f_{BA}^s}\simeq 193.
 \end{equation}
Compared to Eq.~(\ref{exp:first}), it's obvious that they ignored the bidirectional transfer of the primer terminus between the polymerase site and exonuclease site. By our theory, it can be modified as
\begin{equation}
 \phi_e\simeq 1+\frac{\hat{r}_{AB}^s}{f_{BA}^s}=1+\frac{k_{AB}^{se}\sigma}{f_{BA}^s}\simeq 107.
\end{equation}

Here $\sigma=r^e/(r^e+k_{AB}^{es}) = 0.56$, not far from its upper limit at which K.A.Johnson et al.'s expression is recovered. Notice that $\sigma$ could play a negative role if $\sigma\ll 1$ (i.e., $r^e\ll k_{AB}^{es}$), $\sigma = 0.56$ implies that the excision process is highly efficiently employed by T7 DNAP for the proofreading purpose.

\begin{figure}
\includegraphics[width=.6\textwidth]{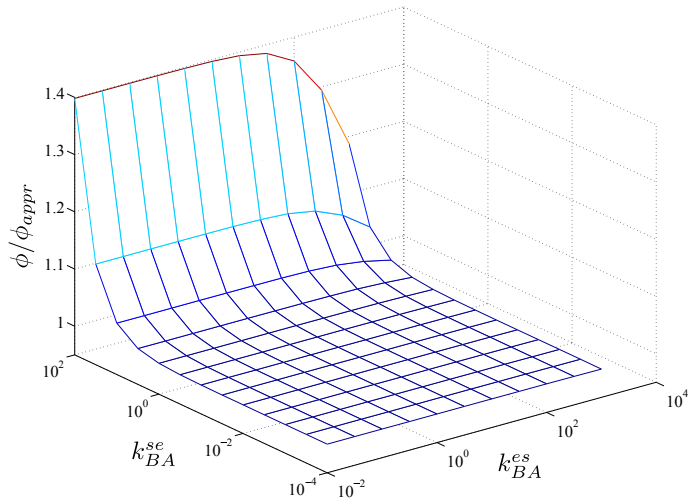}
\caption{The fidelity $\phi$ of T7 DNAP, which is calculated by the exact numerical solution, divided by the approximate expression $\phi_{appr}$. It shows that $\phi$ can be well approximated by $\phi_{appr}$ in large range of the two undetermined parameters $k_{BA}^{se}$ and $k_{BA}^{es}$.
Kinetic parameters are taken from TABLE~\ref{table1} (in unit $s^{-1}$): $f_{AA}^s=250$, $f_{AB}^s=0.002$, $f_{BA}^s=0.012$, $k_{AA}^{se}=0.2$, $k_{AB}^{se}=2.3$, $k_{AA}^{es}=714$, $k_{AB}^{es}=714$, $r^e=896$. All other parameters involving $BB$ are set as zero.}
\label{appr}
\end{figure}

To further validate the approximate expression Eq.~(\ref{exp:first}) for T7 DNAP, we compared the approximate result $\phi_{appr}$ to the exact numerical solution of Eq.~(\ref{1steqs}) $\phi$ in a large range of the two undetermined parameters $k_{BA}^{se}$ and $k_{BA}^{es}$. As shown in FIG.~\ref{appr}, both methods give very close results in large range of $k_{BA}^{se}$ and $k_{BA}^{es}$.

\subsection{Second-order proofreading}

For human mitochondrial DNAP pol $\gamma$, K.A.Johnson et al. measured some kinetic parameters\cite{johnson2001exonuclease} (TABLE~\ref{table1}) which displays the second-order terminal effect.
Although some involved parameters have not been determined directly, some of their combinations were measured. For instance, $\hat{r}_{AAB}^s(=k_{AAB}^{se}r^e/(r^e+k_{AAB}^{es}))=0.4 s^{-1}$ (i.e., $k_{exo}=0.4 s^{-1}$ in Scheme 1 of Ref.~\onlinecite{johnson2001exonuclease}), $\hat{r}_{ABA}^s(=k_{ABA}^{se}r^e/(r^e+k_{ABA}^{es}))=3 s^{-1}$ (i.e., $k_{exo}=3 s^{-1}$ in Scheme 1 of Ref.~\onlinecite{johnson2001exonuclease}).

Assuming that above-mentioned bio-relevant conditions are satisfied and using the available kinetic parameters, one can make a rough estimate of the overall fidelity $\phi=\phi_s \phi_e$ as follows:
\begin{subequations}
\begin{eqnarray}
&&\phi_s\simeq \frac{f_{AAA}^s}{f_{AAB}^s}=\frac{3900-5700}{0.023-1.6} \simeq 10^4-10^5, \\
&&\phi_e\simeq (1+\frac{\hat{r}_{AAB}^{s}}{f_{ABA}^s})(1+\frac{\hat{r}_{ABA}^{s}}{f_{BAA}^s})=(1+\frac{0.4}{0.1})(1+\frac{3}{2.7})\simeq 10.
\end{eqnarray}
\end{subequations}

In their article \cite{johnson2001exonuclease}, K.A.Johnson et al. divided $\phi_e$ intuitively into two multiplying parts. One is due to the correction of the terminal mismatch, and the other is due to the correction of the buried mismatch. In our terminology, they actually considered $(\hat{r}_{AAB}^{s}/f_{ABA}^s)$ and $(\hat{r}_{ABA}^s/f_{BAA}^s)$, respectively.  So their expression of $\phi_e$ is almost the same as ours Eq.~(\ref{exp:second}).

One may notice that the second-order proofreading contribution $(\hat{r}_{ABA}^s/f_{BAA}^s)$, seems insignificant. Fortunately, it can be enhanced, when free dNTP matching the terminal or penultimate base on the template are presented in the solution.
Actually, these dNTPs were observed to apparently accelerate the excision of the penultimate mismatch (see Scheme 2 of Ref.~\onlinecite{johnson2001exonuclease}).
This can be understood by the above expression of $\phi_e$.
Since the duplex terminus is unstable due to the buried mismatch, the free dNTP has the chance to bind transiently to the template at the polymerase site, which may accelerate the transfer of the primer terminus from the polymerase site to the exonuclease site, or hinder the back transfer. This will increase $k_{ABA}^{se}$ or decrease $k_{ABA}^{es}$ (in either case, to increase $\hat{r}_{ABA}^s$), and thus enhance $\phi_e$. In fact, $\hat{r}_{ABA}^s$ was found to increase from 3 $s^{-1}$ (in the absence of matching dNTP in the solution) to up to $21\sim 39 s^{-1}$ (in the presence of matching dNTP), which leads to an order of magnitudes increase of $\phi_e$.

\section{Discussion and conclusion}
In this work, we propose a general kinetic framework to analyze the fidelity problem of DNAP which owns both a polymerase site for primer growth and an exonuclease site for proofreading. So far as we know, it's the first time that the two sub-processes, as well as the higher-order terminal effects, can be rigorously studied in a unified way (either for Model I or for Model II).
Closed equations were derived which fully describe the steady-state replication process.
By these equations, the replication fidelity $\phi$, as well as other quantities such as the total flux $J$ (the overall reaction velocity), can be calculated.
In particular, using the infinite-state Markov chain method which is numerically equivalent to our steady-state equations, we derived analytical expressions of $\phi$ for both Model I and Model II under bio-relevant conditions.
We found that Model I and Model II behave almost the same in every aspect (e.g., the fidelity, the overall reaction velocity, etc.) under those conditions.
This implies that the proofreading efficiency of DNAP may not depend on the details of how the excised primer terminus returns from the exonuclease site to the polymerase site.
Furthermore, the highly simplified expressions of $\phi$ show that the replication fidelity is only determined by very few kinetic parameters, which indicates that the polymerization-proofreading mechanism is insensitive to details of the reaction schemes.

The expression of $\phi$ of $h^{th}$-order model (Eq.~(\ref{eq:fid_tot})) offers intuitive and important insights to understand the higher-order terminal effects.
We noticed that the polymerase site can add $A$ to the primer terminus with a much larger rate than adding $B$, which contributes significantly to the overall fidelity.
In this pathway, however, the $h^{th}$-order terminal effects are not reflected explicitly in $\phi_s$.
In fact, the higher-order effects work in the proofreading pathway.

To simply put, when the primer terminus contains one $B$ at whatever position, it can be extended one $A$ by the polymerase site, or be transferred and excised by the exonuclease site.
Once the former is much larger than the latter (see condition (c), for $0<i \le m$), it can substantially contribute to $\phi$ as a ratio between these two rates.
In principle, for each possible position (the terminal, the penultimate, etc.) of $B$, there is a corresponding ratio contributing to $\phi_e$.
However, it seems only a few leading ratios contribute significantly to $\phi_e$.
As pointed out in Ref.~\onlinecite{petruska1988comparison}, the higher-order effects may originate mainly from base-stacking interaction in the DNA duplex.
The presence of terminal or penultimate mismatch may significantly disrupt the base stacking of the duplex terminus, and thus increases the transfer-and-excision rate and decreases the addition rate, which enhances the
proofreading contribution to the overall fidelity.
On the other hand, deeper mismatches may put less impact on both rates and thus on the proofreading efficiency (see condition (c), for $m < i \le h$).
For instance, in the case of human mitochondrial DNAP pol $\gamma$, it has been observed $f_{BAA}^s \gg f_{ABA}^s$ (TABLE~\ref{table1}), and thus the contribution of the buried mismatch (in the absence of matching dNTP in the solution) to $\phi_e$, $(\hat{r}_{ABA}^s/f_{BAA}^s)$, is smaller than that of the terminal mismatch $(\hat{r}_{AAB}^{s}/f_{ABA}^s)$.
This raises the question that whether the third-order even higher-order effects can be observed for any real DNAPs.
For the third-order model, $\phi_e \simeq \left( 1+r_{AAAB}^s/f_{AABA}^s\right) \left( 1+r_{AABA}^s/f_{ABAA}^s\right) \left( 1+r_{ABAA}^s/f_{BAAA}^s\right)$.
If $f_{BAAA}^s$ approaches to $f_{AAAA}^s$ which is much larger than any other kinetic parameter, then the term corresponding to the third-order effect $r_{ABAA}^s/f_{BAAA}^s$ is negligible, meaning that this has no practical contributions to fidelity.
Whether such higher-order effects exist for other DNAPs is worthy investigation in the future.

It should also be pointed out that we have not discuss the sequence effect on the fidelity in this paper. As shown by experiments such as Ref.~\onlinecite{johnson2001exonuclease}, the 16 possible base pairs may have different incorporation rates or excision rates, which of course have more or less impact on the overall fidelity.
Our model and indeed all the existing models, are actually based on the presumption that the rates of the 4 matches are of similar order of magnitude, and the rates of the 12 mismatches are also of similar order of magnitude.
This coarse-grained description of DNA replication process is appropriate for the purpose to estimate the overall replication fidelity, but cannot account for much subtle effects such as sequence-dependent replication errors which are biologically very important.
To develop a new kinetic theory to take account of the sequence effects would be very challenging, since the symbolic sequence of the template is inevitably involved, which leads to many difficulties for theoretical studies (e.g., it's hard to rigorously define the steady state).
Actually, there have been some numerical simulations in that respect (e.g., see Ref.~\onlinecite{gaspard2014kinetics}), but rigorous modeling of the simulated processes is still lacking.
The approach presented in this paper may serve as a start for further development of such a modeling framework.

\begin{acknowledgments}
The authors thank the financial support by the National Basic Research Program of China (973 program, No.2013CB932804) and National Natural Science Foundation of China (No.11105218, No.91027046, No.11574329 and No.11322543).
\end{acknowledgments}

\appendix

\section{The iterative equation} \label{app:iter}

In the main text, we use $P_{euX_2X_1}$ to denote the possibility that the newly incorporated $X_1$ ever being excised. It can be calculated by counting all the possible routes that lead to the finally excision.

(a) The possibility that the terminal $X_1$ is excised without subsequent dNTP addition can be calculated as $P_{euX_2X_1}^{00}=P_{X_2X_1}^{se} T_{X_2X_1}P_{u|X_2X_1}=\hat{P}_{u|X_2X_1}$. Here $T_{X_2X_1}=1+P_{X_2X_1}^{es}P_{X_2X_1}^{se}+(P_{X_2X_1}^{es}P_{X_2X_1}^{se})^2+\cdots=1/(1-P_{X_2X_1}^{se}P_{X_2X_1}^{es})$. $P_{X_2X_1}^{se} T_{X_2X_1}$ is the sum of the possibility of all the routes that the primer terminus is initially in the polymerase site and then transferred back-and-forth and eventually located at the exonuclease site.

(b) It's also possible that the primer terminal $X_1$ is buried by the next dNTP addition and eventually be excised. For example, $X_1$ is buried by the subsequent addition of $A$ (with a possibility $T_{X_2X_1}P_{A|X_2X_1}$), and this newly added $A$ is excised (with a possibility $P_{euX_1A}$), and finally $X_1$ itself is excised (with a possibility $T_{X_2X_1}P_{u|X_2X_1}$). According to this logic, the possibility of the route that $A$ is incorporated and excised $i$ times and $B$ is incorporated and excised $j$ times before the final excision of $X_1$, can be calculated as $P_{euX_2X_1}^{ij}=C_{i+j}^{i} (T_{X_2X_1} P_{A|X_2X_1}P_{euX_1A})^i (P_{X_2X_1}^{es})^{i+j-1} (T_{X_2X_1} P_{B|X_2X_1}P_{euX_1B})^j (T_{X_2X_1}P_{u|X_2X_1})$ $(i+j\ge 1)$.

Accordingly, we have
\begin{eqnarray}
P_{euX_2X_1} & = & \sum\limits_{i,j\ge 0} P_{euX_2X_1}^{ij} \nonumber \\
& = & \hat{P}_{u|X_2X_1} + \frac{T_{X_2X_1}P_{u|X_2X_1}}{P_{X_2X_1}^{es}}\sum\limits_{i+j\ge 1} C_{i+j}^{i}(\hat{P}_{A|X_2X_1}P_{euX_1A} P_{X_2X_1}^{es})^{i} (\hat{P}_{A|X_2X_1}P_{euX_1B} P_{X_2X_1}^{es})^{j} \nonumber \\
 & = & \hat{P}_{u|X_2X_1} + \frac{\hat{P}_{u|X_2X_1}}{P_{X_2X_1}^{se}P_{X_2X_1}^{es}} \sum\limits_{n\ge 1}(\hat{P}_{A|X_2X_1}P_{euX_1A} P_{X_2X_1}^{es}+\hat{P}_{B|X_2X_1}P_{euX_1B} P_{X_2X_1}^{es})^n \nonumber \\
& = & \hat{P}_{u|X_2X_1} + \frac{\hat{P}_{u|X_2X_1}}{P_{X_2X_1}^{se}P_{X_2X_1}^{es}}(\frac{1}{1-(\hat{P}_{A|X_2X_1}P_{euX_1A} P_{X_2X_1}^{es}+\hat{P}_{B|X_2X_1}P_{euX_1B} P_{X_2X_1}^{es})}-1) \nonumber \\
& = & \frac{\hat{P}_{u|X_2X_1}}{P_{X_2X_1}^{se}P_{X_2X_1}^{es}}(\frac{1}{1-(\hat{P}_{A|X_2X_1}P_{euX_1A}+\hat{P}_{B|X_2X_1}P_{euX_1B}) P_{X_2X_1}^{es}}-\frac{1}{T_{X_2X_1}}).
\end{eqnarray}
For higher-order terminal models, one can also obtain recursion equations of the same form.

\section{The approximation of $\phi$ under bio-relevant conditions} \label{app:approx}

We use the second-order model to demonstrate the approximation.

Under bio-relevant conditions, the fidelity expression can be approximated as,
\begin{eqnarray}
\phi &= & \frac{Q_{AAA} + Q_{ABA} + Q_{BAA}  + Q_{BBA} }{Q_{AAB} + Q_{ABB} +Q_{BAB} + Q_{BBB}}\nonumber \\
 &\simeq & \frac{Q_{AAA} +Q_{ABA} + Q_{BAA} }{Q_{AAB} } \\ \nonumber
 &\simeq & \frac{Q_{AAA}}{Q_{AAB}} + 2  \simeq \frac{\hat{P}_{AAA}P_{nuAAA}}{\hat{P}_{AAB}P_{nuAAB}}.
\end{eqnarray}

In the first step, we have $Q_{ABB}= Q_{BAB} =Q_{BBA} =Q_{BBB}=0$ because of condition (d). In the second step, we have $Q_{AAA} \gg Q_{AAB} $ because of the conditions (a) and (b), and $Q_{ABA}=Q_{BAA}=Q_{AAB} $ due to constraint Eq.~(\ref{eq:r2}) (i.e., $Q_{XY}=Q_{AXY}+Q_{BXY}=Q_{XYA}+Q_{XYB}$).

The fidelity expression can then be separated into two parts, $\phi_s=\hat{P}_{AAA}/\hat{P}_{AAB}$ and $\phi_e=P_{nuAAA}/P_{nuAAB}$. The first part is the contribution of polymerase site, which can be easily calculated as $\phi_s=\hat{P}_{AAA}/\hat{P}_{AAB}=f_{AAA}^s/f_{AAB}^s$. The second part $\phi_e$ is the contribution of exonuclease site, which can be calculated as follows.

First, $P_{nuAAA}=1-P_{euAAA}\simeq 1$, since $P_{euAAA}\simeq 0$ (this is intuitive according to conditions (a) and (b), and can be verified by numerical calculation). Thus, the fidelity $\phi_e$ is determined by $P_{nuAAB}=1-P_{euAAB}$. For $P_{euAAB}$, similar to Appendix \ref{app:iter}, we have
\begin{eqnarray}
P_{euAAB} & = & \frac{\hat{P}_{u|AAB}}{P_{AAB}^{se}P_{AAB}^{es}}(\frac{1}{1-(\hat{P}_{A|AAB}P_{euABA}+\hat{P}_{B|AAB}P_{euABB})P_{AAB}^{es}}-\frac{1}{T_{AAB}}) \nonumber \\
& = & \frac{\hat{P}_{u|AAB}}{P_{AAB}^{se}P_{AAB}^{es}}(\frac{1}{1-(\mathcal{A}_1^{(2)}+\mathcal{B}_1^{(2)})P_{AAB}^{es}}-\frac{1}{T_{AAB}}) \\ \nonumber
& = & \frac{\hat{P}_{u|AAB}}{P_{AAB}^{se}P_{AAB}^{es}}(\frac{1}{1-\mathcal{A}_1^{(2)}P_{AAB}^{es}}-\frac{1}{T_{AAB}}),
\end{eqnarray}
where $\mathcal{A}_1^{(2)}=\hat{P}_{A|AAB}P_{euABA}$, $\mathcal{B}_1^{(2)}=\hat{P}_{B|AAB}P_{euABB}$. In the second step, we used $\mathcal{B}_1^{(2)}=0$ because of condition (d). Now all the quantities in the expression of $P_{euAAB}$ are known except $P_{euABA}$ which can be expressed as,
\begin{eqnarray}
P_{euABA} & = & \frac{\hat{P}_{u|ABA}}{P_{ABA}^{se}P_{ABA}^{es}}(\frac{1}{1-(\hat{P}_{A|ABA}P_{euBAA}+\hat{P}_{B|ABA}P_{euBAB})P_{ABA}^{es}}-\frac{1}{T_{ABA}}) \nonumber \\
& = & \frac{\hat{P}_{u|ABA}}{P_{ABA}^{se}P_{ABA}^{es}}(\frac{1}{1-(\mathcal{A}_2^{(2)}+\mathcal{B}_2^{(2)})P_{ABA}^{es}}-\frac{1}{T_{ABA}}) \\ \nonumber
& = & \frac{\hat{P}_{u|ABA}}{P_{ABA}^{se}P_{ABA}^{es}}(\frac{1}{1-\mathcal{A}_2^{(2)}P_{ABA}^{es}}-\frac{1}{T_{ABA}}),
\end{eqnarray}
where $\mathcal{A}_2^{(2)}=\hat{P}_{A|ABA}P_{euBAA}$, $\mathcal{B}_2^{(2)}=\hat{P}_{B|ABA}P_{euBAB}$. In the second step, we used $\mathcal{B}_2^{(2)}=0$ because of condition (d). As for $\mathcal{A}_2^{(2)}$, it's actually negligible. To make it clear, we resort to the expression of $P_{euBAA}$:
\begin{eqnarray}
P_{euBAA} & = & \frac{\hat{P}_{u|BAA}}{P_{BAA}^{se}P_{BAA}^{es}}(\frac{1}{1-(\hat{P}_{A|BAA}P_{euAAA}+\hat{P}_{B|BAA}P_{euAAB})P_{BAA}^{es}}-\frac{1}{T_{BAA}}) \nonumber \\
& = & \frac{\hat{P}_{u|BAA}}{P_{BAA}^{se}P_{BAA}^{es}}(\frac{1}{1-(\mathcal{A}_3^{(2)}+\mathcal{B}_3^{(2)})P_{BAA}^{es}}-\frac{1}{T_{BAA}}) \nonumber \\
& = & \hat{P}_{u|BAA}  \simeq 0,
\end{eqnarray}
where $\mathcal{A}_3^{(2)}=\hat{P}_{A|BAA}P_{euAAA}$, $\mathcal{B}_3^{(2)}=\hat{P}_{B|BAA}P_{euAAB}$. $\mathcal{A}_3^{(2)}\simeq 0$ since $P_{euAAA}=0$.
$\mathcal{B}_3^{(2)}\simeq 0$ since $\hat{P}_{B|BAA} \simeq 0$ because of condition (a).
Finally, we obtain $P_{euBAA}\simeq \hat{P}_{u|BAA}\simeq 0$
because of condition (a) and (b).

Now we have $P_{euABA}\simeq \hat{P}_{u|ABA}$, $\mathcal{A}_1^{(2)}=\hat{P}_{A|AAB}\hat{P}_{u|ABA}$, and
\begin{eqnarray}
\label{eq:better}
P_{euAAB}  \simeq  \frac{\hat{P}_{u|AAB}}{P_{AAB}^{se}P_{AAB}^{es}}(\frac{1}{1-\hat{P}_{A|AAB}\hat{P}_{u|ABA}P_{AAB}^{es}}-\frac{1}{T_{AAB}}).
\end{eqnarray}

This expression is of the following general form:
\begin{equation}\label{m1_p1}
p_1=\frac{\alpha}{\theta\gamma}(\frac{1}{1-(1-\alpha)\beta\gamma}-(1-\theta\gamma)),
\end{equation}
where $\alpha=\theta(1-\gamma)/(1-\theta\gamma)$ and $0 <\alpha, \beta, \gamma, \theta< 1$.
It can be approximated by the following simpler expression:
\begin{equation}\label{m1_p2}
p_2= \alpha+(1-\alpha)\beta.
\end{equation}

\begin{figure}
\includegraphics[width=0.97\textwidth]{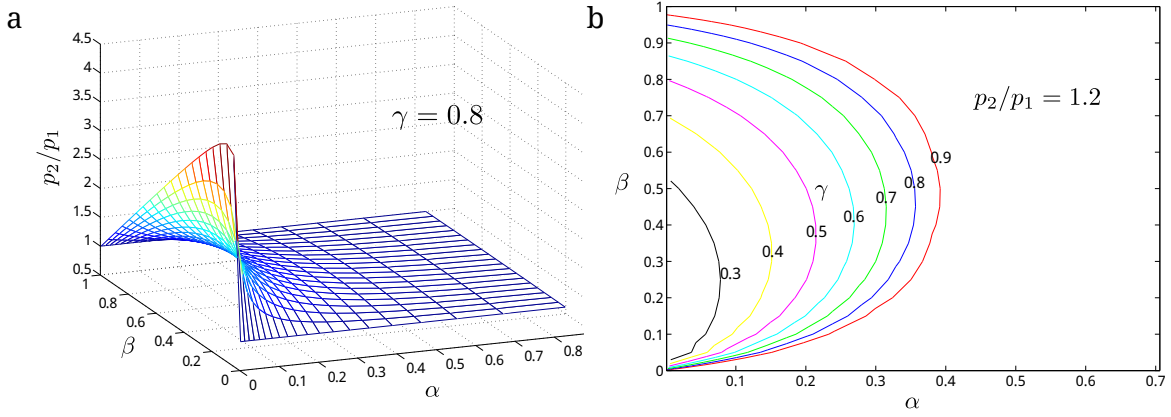}
\caption{(a) Comparison between $p_2$ and $p_1$ in the parameter space ($\alpha, \beta$). $\gamma$ is set to $0.8$. (b) The isolines of $p_2/p_1=1.2$, for $\gamma$ varying from $0.3$ to $0.9$. Right to all isolines is the range of $\alpha$ and $\beta$ in which $p_2/p_1$ almost equals to 1.}
\label{model1_comp}
\end{figure}

$p_1\simeq p_2$ holds for $\alpha \agt 0.5$ (here it means $\hat{P}_{u|AAB} \agt 0.5$, i.e., $R_i\agt F_i$ for $0<i\le 1$, see condition (c)), which can be verified numerically as shown by FIG.~\ref{model1_comp}. Thus, we can write $P_{euAAB}$ as,
\begin{eqnarray}\label{better1}
P_{euAAB} \simeq  \hat{P}_{u|AAB}+ \hat{P}_{A|AAB}\hat{P}_{u|ABA}.
\end{eqnarray}

The fidelity $\phi_e\simeq 1/(1-P_{euAAB})$ can then be calculated. Finally we obtain an intuitive approximate expression of the overall replication fidelity:
\begin{equation} \label{exp:fid2}
\phi\simeq \frac{\hat{P}_{AAA}}{\hat{P}_{AAB}(1-P_{euAAB})}=\frac{\hat{P}_{AAA}}{\hat{P}_{AAB}\hat{P}_{A|AAB}\hat{P}_{A|ABA}}=
\frac{f_{AAA}^s}{f_{AAB}^s}(1+\frac{\hat{r}_{AAB}^s}{f_{ABA}^s})(1+\frac{\hat{r}_{ABA}^s}{f_{BAA}^s}).
\end{equation}

It should be noted that under conditions that $\alpha \ll 1$ and $\beta \ll 1$(i.e., $R_i\ll F_i$ for $0< i\le 2$, see condition (c)), the expression Eq.~(\ref{exp:fid2}) is still valid, since $p_1\simeq 0$ and thus $\phi_e\simeq 1$.

Extending this logic to higher-order models is straightforward. For $h^{th}$-order model, we similarly have
\begin{eqnarray}
\phi &&\equiv \frac{{\sum\limits_{X_i=A,B}}\hat{P}_{X_{h+1}X_h\cdots X_2A}P_{nuX_{h+1}X_h\cdots X_2A}}{\sum\limits_{X_i=A,B}\hat{P}_{X_{h+1}X_h\cdots X_2B}P_{nuX_{h+1}X_h\cdots X_2B}}\simeq \frac{\hat{P}_{\underbrace{\scriptscriptstyle{AAA}\cdots}_{h+1}}P_{nu\underbrace{\scriptscriptstyle{AAA}\cdots}_{h+1}}}
{\hat{P}_{\underbrace{\scriptscriptstyle{AAA}\cdots}_{h}B}P_{nu\underbrace{\scriptscriptstyle{AAA}\cdots}_{h}B}}+h \nonumber \\
&&\simeq \frac{\hat{P}_{\underbrace{\scriptscriptstyle{AAA}\cdots}_{h+1}}P_{nu\underbrace{\scriptscriptstyle{AAA}\cdots}_{h+1}}}{\hat{P}_{\underbrace{\scriptscriptstyle{AAA}\cdots}_{h}B}P_{nu\underbrace{\scriptscriptstyle{AAA}\cdots}_{h}B}},
\end{eqnarray}
and we also have $P_{eu\underbrace{\scriptscriptstyle{AAA}\cdots}_{h+1}}\simeq 0$, thus $P_{nu\underbrace{\scriptscriptstyle{AAA}\cdots}_{h+1}}\simeq 1$. To calculate $P_{nu\underbrace{\scriptscriptstyle{AAA}\cdots}_{h}B}=1-P_{eu\underbrace{\scriptscriptstyle{AAA}\cdots}_{h}B}$, we have to calculate all the following,
\begin{eqnarray*}
R_i^\ast\equiv P_{eu\underbrace{\scriptscriptstyle{AAA}\cdots}_{h-i+1}B\underbrace{\scriptscriptstyle{AAA}\cdots}_{i-1}}\simeq \frac{\hat{P}_{u|\underbrace{\scriptscriptstyle{AAA}\cdots}_{h-i+1}B\underbrace{\scriptscriptstyle{AAA}\cdots}_{i-1}}}{P_{\underbrace{\scriptscriptstyle{AAA}\cdots}_{h-i+1}B\underbrace{\scriptscriptstyle{AAA}\cdots}_{i-1}}^{se}P_{\underbrace{\scriptscriptstyle{AAA}\cdots}_{h-i+1}B\underbrace{\scriptscriptstyle{AAA}\cdots}_{i-1}}^{es}}(\frac{1}{1-(\mathcal{A}_{i}^{(h)}+\mathcal{B}_{i}^{(h)})P_{\underbrace{\scriptscriptstyle{AAA}\cdots}_{h-i+1}B\underbrace{\scriptscriptstyle{AAA}\cdots}_{i-1}}^{es}}-\frac{1}{T_{\underbrace{\scriptscriptstyle{AAA}\cdots}_{h-i+1}B\underbrace{\scriptscriptstyle{AAA}\cdots}_{i-1}}}),
\end{eqnarray*}
$1\le i \le h+1$, where $\mathcal{A}_{i}^{(h)}\equiv \hat{P}_{A|\underbrace{\scriptscriptstyle{AAA}\cdots}_{h-i+1}B\underbrace{\scriptscriptstyle{AAA}\cdots}_{i-1}}
P_{eu\underbrace{\scriptscriptstyle{AAA}\cdots}_{h-i}B\underbrace{\scriptscriptstyle{AAA}\cdots}_{i}}$, $\mathcal{B}_{i}^{(h)}\equiv \hat{P}_{B|\underbrace{\scriptscriptstyle{AAA}\cdots}_{h-i+1}B\underbrace{\scriptscriptstyle{AAA}\cdots}_{i-1}}
P_{eu\underbrace{\scriptscriptstyle{AAA}\cdots}_{h-i}B\underbrace{\scriptscriptstyle{AAA}\cdots}_{i-1}B}$ for $1\le i \le h$, and $\mathcal{A}_{h+1}^{(h)}\equiv \hat{P}_{A|B\underbrace{\scriptscriptstyle{AAA}\cdots}_{h}}
P_{eu\underbrace{\scriptscriptstyle{AAA}\cdots}_{h+1}}$, $\mathcal{B}_{h+1}^{(h)}\equiv \hat{P}_{B|B\underbrace{\scriptscriptstyle{AAA}\cdots}_{h}}
P_{eu\underbrace{\scriptscriptstyle{AAA}\cdots}_{h}B}$.
We have $\mathcal{B}_{i}^{(h)} \simeq 0 (1\le i\le h)$ because of condition (d), $\mathcal{B}_{h+1}^{(h)} \simeq 0$ because of condition (a), and $\mathcal{A}_{h+1}^{(h)} \simeq 0$ since $P_{eu\underbrace{\scriptscriptstyle{AAA}\cdots}_{h+1}}\simeq 0$. So we obtain $R_{h+1}^\ast \simeq \hat{P}_{u|B\underbrace{\scriptscriptstyle{AAA}\cdots}_{h}} \simeq 0$ because of condition (a) and (b).

As defined in the main text, $F_i\equiv \hat{P}_{A|\underbrace{\scriptscriptstyle{AAA}\cdots}_{h-i+1}B\underbrace{\scriptscriptstyle{AAA}\cdots}_{i-1}}$, and 
$R_i \equiv \hat{P}_{u|\underbrace{\scriptscriptstyle{AAA}\cdots}_{h-i+1}B\underbrace{\scriptscriptstyle{AAA}\cdots}_{i-1}}$.
For $1\le i\le h$, they can be written as (due to condition (d)):
\begin{subequations}
\begin{eqnarray}
F_i= \frac{f_{\underbrace{\scriptscriptstyle{AAA}\cdots}_{h-i}B\underbrace{\scriptscriptstyle{AAA}\cdots}_{i}}^s}
{f_{\underbrace{\scriptscriptstyle{AAA}\cdots}_{h-i}B\underbrace{\scriptscriptstyle{AAA}\cdots}_{i}}^s+
\hat{r}_{\underbrace{\scriptscriptstyle{AAA}\cdots}_{h-i+1}B\underbrace{\scriptscriptstyle{AAA}\cdots}_{i-1}}^s}, \\
R_i=\frac{\hat{r}_{\underbrace{\scriptscriptstyle{AAA}\cdots}_{h-i+1}B\underbrace{\scriptscriptstyle{AAA}\cdots}_{i-1}}^s}
{f_{\underbrace{\scriptscriptstyle{AAA}\cdots}_{h-i}B\underbrace{\scriptscriptstyle{AAA}\cdots}_{i}}^s
+\hat{r}_{\underbrace{\scriptscriptstyle{AAA}\cdots}_{s-i+1}B\underbrace{\scriptscriptstyle{AAA}\cdots}_{i-1}}^s},
\end{eqnarray}
\end{subequations}
and obviously $F_i+R_i=1$.

To calculate other $R_i^\ast$, we first notice that $\mathcal{A}_{h}^{(h)} \simeq 0$ and thus $R_{h}^\ast \simeq R_h$, since $R_{h+1}^\ast \simeq 0$. For $R_i^\ast$ ($1\le i \le h-1$), we can express it in a form similar to Eq.~(\ref{m1_p1}) and thus it can be rewritten as Eq.~(\ref{m1_p2}) under conditions (c$^\prime$) $R_i \agt F_i$($1\le i \le h-1$):
\begin{eqnarray}
R_i^{\ast}=R_i+F_iR_{i+1}^\ast (1\le i \le h).
\end{eqnarray}

Hence, it can be easily proved that
\begin{eqnarray}
\phi_e \simeq \frac{1}{1-R_1^{\ast}} \simeq \frac{1}{F_1}\frac{1}{F_2}\ldots\frac{1}{F_h}= (1+\frac{R_1}{F_1})(1+\frac{R_2}{F_2})(\ldots)(1+\frac{R_h}{F_h}).
\end{eqnarray}

Thus, for $h^{th}$-order model, we have,
\begin{eqnarray} \label{app:higher}
\phi=\phi_s\phi_e \simeq \frac{f_{\underbrace{\scriptscriptstyle{AAA\cdots}}_hA}^s}{f_{\underbrace{\scriptscriptstyle{AAA\cdots}}_hB}^s}
(1+\frac{R_1}{F_1})(1+\frac{R_2}{F_2})(\cdots)(1+\frac{R_h}{F_h}).
\end{eqnarray}

Similar to the second-order model, it can be seen that this expression is still valid under the condition (c) which is less restrictive and more practical than condition (c$^\prime$).

\section{Approximation of $\phi$ for Model II} \label{app:m2}

\begin{figure}
\includegraphics[width=.5\textwidth]{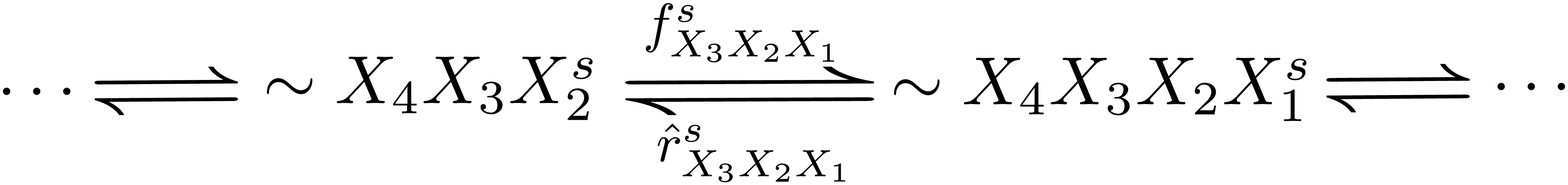}
\caption{The minimal second-order reaction scheme for Model II}
\label{model2b}
\end{figure}

The minimal second-order scheme of Model II is shown as FIG.~\ref{model2b}. The effective excision rate $\hat{r}_{X_3X_2X_1}$ is the same as that in Model I, which is $\hat{r}_{X_3X_2X_1}^s=k_{X_3X_2X_1}^{se}r_{X_3X_2X_1}^e/(r_{X_3X_2X_1}^e+k_{X_3X_2X_1}^{es})$.
As shown in the supplement of our previous paper\cite{shu2015general}, infinite-state Markov chain method can also apply to Model II, and the iterative expression for $P_{euX_3X_2X_1}$ is,
\begin{equation}
P_{euX_3X_2X_1}  = \frac{\hat{P}_{u|X_3X_2X_1}}{1-(\hat{P}_{A|X_3X_2X_1}P_{euX_2X_1A}+\hat{P}_{B|X_3X_2X_1}P_{euX_2X_1B})}.
\end{equation}

We also define
$\hat{P}_{u|X_3X_2X_1}=\hat{r}_{X_3X_2X_1}^s/(\hat{r}_{X_3X_2X_1}^s+f_{X_2X_1A}^s+f_{X_2X_1B}^s)$, and $\hat{P}_{X|X_3X_2X_1}=f_{X_2X_1X}^s/(\hat{r}_{X_3X_2X_1}^s+f_{X_2X_1X}^s+f_{X_2X_1\bar{X}}^s)$.

Under bio-relevant conditions, the fidelity expression can be approximated as,
\begin{eqnarray}
\phi &= & \frac{Q_{AAA} + Q_{ABA} + Q_{BAA}  + Q_{BBA} }{Q_{AAB} + Q_{ABB} +Q_{BAB} + Q_{BBB}} \nonumber \\
 &\simeq & \frac{\hat{P}_{AAA}P_{nuAAA}}{\hat{P}_{AAB}P_{nuAAB}} \\
 & = & \phi_s\phi_e, \nonumber
\end{eqnarray}
where $\phi_s=\hat{P}_{AAA}/\hat{P}_{AAB}=f_{AAA}^s/f_{AAB}^s$, $\phi_e=P_{nuAAA}/P_{nuAAB}\simeq 1/P_{nuAAB}$.

Therefore, the fidelity $\phi_e$ is determined by $P_{nuAAB}=1-P_{euAAB}$. For $P_{euAAB}$,
\begin{eqnarray}
P_{euAAB}& = & \frac{\hat{P}_{u|AAB}}{1-(\hat{P}_{A|AAB}P_{euABA}+\hat{P}_{B|AAB}P_{euABB})} \nonumber \\
\label{eq:m2_aab}
& \simeq & \frac{\hat{P}_{u|AAB}}{1-\hat{P}_{A|AAB}P_{euABA}}.
\end{eqnarray}

$P_{euABA}$ can be calculated as,
\begin{eqnarray}
P_{euABA}& = & \frac{\hat{P}_{u|ABA}}{1-(\hat{P}_{A|ABA}P_{euBAA}+\hat{P}_{B|ABA}P_{euBAB})} \nonumber\\
& \simeq & \frac{\hat{P}_{u|ABA}}{1-\hat{P}_{A|ABA}P_{euBAA}}.
\end{eqnarray}

$P_{euBAA}$ is shown to be negligible:
\begin{eqnarray}
P_{euBAA}& = & \frac{\hat{P}_{u|BAA}}{1-(\hat{P}_{A|BAA}P_{euAAA}+\hat{P}_{B|BAA}P_{euAAB})} \nonumber \\
& \simeq & \frac{\hat{P}_{u|BAA}}{1-\hat{P}_{A|BAA}P_{euAAA}} \\
& \simeq & \hat{P}_{u|BAA}  \simeq  0. \nonumber
\end{eqnarray}

So we have $P_{euABA}\simeq \hat{P}_{u|ABA}$. Substituting it to Eq.~(\ref{eq:m2_aab}), we obtain
\begin{equation}
\label{eq:m2_fid1}
P_{euAAB} \simeq  \hat{P}_{u|AAB}(\frac{1}{1-\hat{P}_{A|AAB}\hat{P}_{u|ABA}}),
\end{equation}
which is of the general form:
\begin{equation}
p_1^\prime=\alpha(\frac{1}{1-(1-\alpha)\beta}).
\end{equation}

\begin{figure}
\includegraphics[width=.6\textwidth]{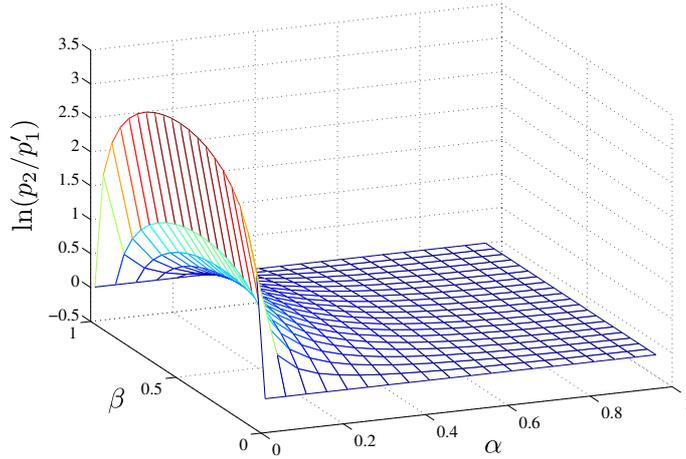}
\caption{Comparison between $p_1^\prime$ and $p_2$, in the parameter space ($\alpha,\beta$).}
\label{model2_comp}
\end{figure}

Numerical calculations show that Eq.~(\ref{m1_p2}) is also a good approximation of such a function under the condition that $\alpha \agt 0.5$, as indicated by FIG.~\ref{model2_comp}. Thus, we obtain the same expression of $P_{euAAB}$ for both Model I and Mode II:
\begin{equation}
\label{eq:m2_fid2}
P_{euAAB}\simeq  \hat{P}_{u|AAB} + \hat{P}_{A|AAB}\hat{P}_{u|ABA}.
\end{equation}
The overall fidelity can be expressed by Eq.~(\ref{exp:fid2}).
When $\hat{P}_{u|AAB}\ll 1$ and $\hat{P}_{u|ABA}\ll 1$, we have $P_{euAAB}\simeq \hat{P}_{u|AAB}\simeq 0$, and the expression is still valid.

Similar to Model I, this result can be directly extended to higher-order models under condition (c$^\prime$) and (c), which gives the same expression as Eq.~(\ref{app:higher}) in Model I.

%\iffalse

\nocite{*}


%\fi


\begin{thebibliography}{51}%
\makeatletter
\providecommand \@ifxundefined [1]{%
 \@ifx{#1\undefined}
}%
\providecommand \@ifnum [1]{%
 \ifnum #1\expandafter \@firstoftwo
 \else \expandafter \@secondoftwo
 \fi
}%
\providecommand \@ifx [1]{%
 \ifx #1\expandafter \@firstoftwo
 \else \expandafter \@secondoftwo
 \fi
}%
\providecommand \natexlab [1]{#1}%
\providecommand \enquote  [1]{``#1''}%
\providecommand \bibnamefont  [1]{#1}%
\providecommand \bibfnamefont [1]{#1}%
\providecommand \citenamefont [1]{#1}%
\providecommand \href@noop [0]{\@secondoftwo}%
\providecommand \href [0]{\begingroup \@sanitize@url \@href}%
\providecommand \@href[1]{\@@startlink{#1}\@@href}%
\providecommand \@@href[1]{\endgroup#1\@@endlink}%
\providecommand \@sanitize@url [0]{\catcode `\\12\catcode `\$12\catcode
  `\&12\catcode `\#12\catcode `\^12\catcode `\_12\catcode `\%12\relax}%
\providecommand \@@startlink[1]{}%
\providecommand \@@endlink[0]{}%
\providecommand \url  [0]{\begingroup\@sanitize@url \@url }%
\providecommand \@url [1]{\endgroup\@href {#1}{\urlprefix }}%
\providecommand \urlprefix  [0]{URL }%
\providecommand \Eprint [0]{\href }%
\providecommand \doibase [0]{http://dx.doi.org/}%
\providecommand \selectlanguage [0]{\@gobble}%
\providecommand \bibinfo  [0]{\@secondoftwo}%
\providecommand \bibfield  [0]{\@secondoftwo}%
\providecommand \translation [1]{[#1]}%
\providecommand \BibitemOpen [0]{}%
\providecommand \bibitemStop [0]{}%
\providecommand \bibitemNoStop [0]{.\EOS\space}%
\providecommand \EOS [0]{\spacefactor3000\relax}%
\providecommand \BibitemShut  [1]{\csname bibitem#1\endcsname}%
\let\auto@bib@innerbib\@empty
%</preamble>
\bibitem [{\citenamefont {Watson}, \citenamefont {Crick}\ \emph
  {et~al.}(1953)\citenamefont {Watson}, \citenamefont {Crick} \emph
  {et~al.}}]{watson1953molecular}%
  \BibitemOpen
  \bibfield  {author} {\bibinfo {author} {\bibfnamefont {J.~D.}\ \bibnamefont
  {Watson}}, \bibinfo {author} {\bibfnamefont {F.~H.}\ \bibnamefont {Crick}},
  \emph {et~al.},\ }\href@noop {} {\bibfield  {journal} {\bibinfo  {journal}
  {Nature}\ }\textbf {\bibinfo {volume} {171}},\ \bibinfo {pages} {737}
  (\bibinfo {year} {1953})}\BibitemShut {NoStop}%
\bibitem [{\citenamefont {Lehman}\ \emph {et~al.}(1958)\citenamefont {Lehman},
  \citenamefont {Bessman}, \citenamefont {Simms},\ and\ \citenamefont
  {Kornberg}}]{lehman1958enzymatic}%
  \BibitemOpen
  \bibfield  {author} {\bibinfo {author} {\bibfnamefont {I.}~\bibnamefont
  {Lehman}}, \bibinfo {author} {\bibfnamefont {M.~J.}\ \bibnamefont {Bessman}},
  \bibinfo {author} {\bibfnamefont {E.~S.}\ \bibnamefont {Simms}}, \ and\
  \bibinfo {author} {\bibfnamefont {A.}~\bibnamefont {Kornberg}},\ }\href@noop
  {} {\bibfield  {journal} {\bibinfo  {journal} {J. biol. Chem}\ }\textbf
  {\bibinfo {volume} {233}},\ \bibinfo {pages} {163} (\bibinfo {year}
  {1958})}\BibitemShut {NoStop}%
\bibitem [{\citenamefont {Kunkel}\ and\ \citenamefont
  {Bebenek}(2000)}]{kunkel2000dna}%
  \BibitemOpen
  \bibfield  {author} {\bibinfo {author} {\bibfnamefont {T.~A.}\ \bibnamefont
  {Kunkel}}\ and\ \bibinfo {author} {\bibfnamefont {K.}~\bibnamefont
  {Bebenek}},\ }\href@noop {} {\bibfield  {journal} {\bibinfo  {journal}
  {Annual review of biochemistry}\ }\textbf {\bibinfo {volume} {69}},\ \bibinfo
  {pages} {497} (\bibinfo {year} {2000})}\BibitemShut {NoStop}%
\bibitem [{\citenamefont {Hopfield}(1974)}]{hopfield1974kinetic}%
  \BibitemOpen
  \bibfield  {author} {\bibinfo {author} {\bibfnamefont {J.~J.}\ \bibnamefont
  {Hopfield}},\ }\href@noop {} {\bibfield  {journal} {\bibinfo  {journal}
  {Proceedings of the National Academy of Sciences}\ }\textbf {\bibinfo
  {volume} {71}},\ \bibinfo {pages} {4135} (\bibinfo {year}
  {1974})}\BibitemShut {NoStop}%
\bibitem [{\citenamefont {Ninio}(1975)}]{ninio1975kinetic}%
  \BibitemOpen
  \bibfield  {author} {\bibinfo {author} {\bibfnamefont {J.}~\bibnamefont
  {Ninio}},\ }\href@noop {} {\bibfield  {journal} {\bibinfo  {journal}
  {Biochimie}\ }\textbf {\bibinfo {volume} {57}},\ \bibinfo {pages} {587}
  (\bibinfo {year} {1975})}\BibitemShut {NoStop}%
\bibitem [{\citenamefont {Galas}\ and\ \citenamefont
  {Branscomb}(1978)}]{galas1978enzymatic}%
  \BibitemOpen
  \bibfield  {author} {\bibinfo {author} {\bibfnamefont {D.~J.}\ \bibnamefont
  {Galas}}\ and\ \bibinfo {author} {\bibfnamefont {E.~W.}\ \bibnamefont
  {Branscomb}},\ }\href@noop {} {\bibfield  {journal} {\bibinfo  {journal}
  {Journal of molecular biology}\ }\textbf {\bibinfo {volume} {124}},\ \bibinfo
  {pages} {653} (\bibinfo {year} {1978})}\BibitemShut {NoStop}%
\bibitem [{\citenamefont {Clayton}\ \emph {et~al.}(1979)\citenamefont
  {Clayton}, \citenamefont {Goodman}, \citenamefont {Branscomb},\ and\
  \citenamefont {Galas}}]{clayton1979error}%
  \BibitemOpen
  \bibfield  {author} {\bibinfo {author} {\bibfnamefont {L.~K.}\ \bibnamefont
  {Clayton}}, \bibinfo {author} {\bibfnamefont {M.~F.}\ \bibnamefont
  {Goodman}}, \bibinfo {author} {\bibfnamefont {E.~W.}\ \bibnamefont
  {Branscomb}}, \ and\ \bibinfo {author} {\bibfnamefont {D.~J.}\ \bibnamefont
  {Galas}},\ }\href@noop {} {\bibfield  {journal} {\bibinfo  {journal} {Journal
  of Biological Chemistry}\ }\textbf {\bibinfo {volume} {254}},\ \bibinfo
  {pages} {1902} (\bibinfo {year} {1979})}\BibitemShut {NoStop}%
\bibitem [{\citenamefont {Johnson}(1993)}]{johnson1993conformational}%
  \BibitemOpen
  \bibfield  {author} {\bibinfo {author} {\bibfnamefont {K.~A.}\ \bibnamefont
  {Johnson}},\ }\href@noop {} {\bibfield  {journal} {\bibinfo  {journal}
  {Annual review of biochemistry}\ }\textbf {\bibinfo {volume} {62}},\ \bibinfo
  {pages} {685} (\bibinfo {year} {1993})}\BibitemShut {NoStop}%
\bibitem [{\citenamefont {Goodman}(1997)}]{goodman1997hydrogen}%
  \BibitemOpen
  \bibfield  {author} {\bibinfo {author} {\bibfnamefont {M.~F.}\ \bibnamefont
  {Goodman}},\ }\href@noop {} {\bibfield  {journal} {\bibinfo  {journal}
  {Proceedings of the National Academy of Sciences}\ }\textbf {\bibinfo
  {volume} {94}},\ \bibinfo {pages} {10493} (\bibinfo {year}
  {1997})}\BibitemShut {NoStop}%
\bibitem [{\citenamefont {Goodman}\ and\ \citenamefont
  {Fygenson}(1998)}]{goodman1998dna}%
  \BibitemOpen
  \bibfield  {author} {\bibinfo {author} {\bibfnamefont {M.~F.}\ \bibnamefont
  {Goodman}}\ and\ \bibinfo {author} {\bibfnamefont {D.~K.}\ \bibnamefont
  {Fygenson}},\ }\href@noop {} {\bibfield  {journal} {\bibinfo  {journal}
  {Genetics}\ }\textbf {\bibinfo {volume} {148}},\ \bibinfo {pages} {1475}
  (\bibinfo {year} {1998})}\BibitemShut {NoStop}%
\bibitem [{\citenamefont {Fersht}(1979)}]{fersht1979fidelity}%
  \BibitemOpen
  \bibfield  {author} {\bibinfo {author} {\bibfnamefont {A.~R.}\ \bibnamefont
  {Fersht}},\ }\href@noop {} {\bibfield  {journal} {\bibinfo  {journal}
  {Proceedings of the National Academy of Sciences}\ }\textbf {\bibinfo
  {volume} {76}},\ \bibinfo {pages} {4946} (\bibinfo {year}
  {1979})}\BibitemShut {NoStop}%
\bibitem [{\citenamefont {Patel}, \citenamefont {Wong},\ and\ \citenamefont
  {Johnson}(1991)}]{patel1991pre}%
  \BibitemOpen
  \bibfield  {author} {\bibinfo {author} {\bibfnamefont {S.~S.}\ \bibnamefont
  {Patel}}, \bibinfo {author} {\bibfnamefont {I.}~\bibnamefont {Wong}}, \ and\
  \bibinfo {author} {\bibfnamefont {K.~A.}\ \bibnamefont {Johnson}},\
  }\href@noop {} {\bibfield  {journal} {\bibinfo  {journal} {Biochemistry}\
  }\textbf {\bibinfo {volume} {30}},\ \bibinfo {pages} {511} (\bibinfo {year}
  {1991})}\BibitemShut {NoStop}%
\bibitem [{\citenamefont {Cline}, \citenamefont {Braman},\ and\ \citenamefont
  {Hogrefe}(1996)}]{Cline1996PCR}%
  \BibitemOpen
  \bibfield  {author} {\bibinfo {author} {\bibfnamefont {J.}~\bibnamefont
  {Cline}}, \bibinfo {author} {\bibfnamefont {J.~C.}\ \bibnamefont {Braman}}, \
  and\ \bibinfo {author} {\bibfnamefont {H.~H.}\ \bibnamefont {Hogrefe}},\
  }\href@noop {} {\bibfield  {journal} {\bibinfo  {journal} {Nucleic Acids
  Research}\ }\textbf {\bibinfo {volume} {24}},\ \bibinfo {pages} {3546}
  (\bibinfo {year} {1996})}\BibitemShut {NoStop}%
\bibitem [{\citenamefont {Wuite}\ \emph {et~al.}(2000)\citenamefont {Wuite},
  \citenamefont {Smith}, \citenamefont {Young}, \citenamefont {Keller},\ and\
  \citenamefont {Bustamante}}]{wuite2000single}%
  \BibitemOpen
  \bibfield  {author} {\bibinfo {author} {\bibfnamefont {G.~J.}\ \bibnamefont
  {Wuite}}, \bibinfo {author} {\bibfnamefont {S.~B.}\ \bibnamefont {Smith}},
  \bibinfo {author} {\bibfnamefont {M.}~\bibnamefont {Young}}, \bibinfo
  {author} {\bibfnamefont {D.}~\bibnamefont {Keller}}, \ and\ \bibinfo {author}
  {\bibfnamefont {C.}~\bibnamefont {Bustamante}},\ }\href@noop {} {\bibfield
  {journal} {\bibinfo  {journal} {Nature}\ }\textbf {\bibinfo {volume} {404}},\
  \bibinfo {pages} {103} (\bibinfo {year} {2000})}\BibitemShut {NoStop}%
\bibitem [{\citenamefont {Tsai}\ and\ \citenamefont
  {Johnson}(2006)}]{tsai2006new}%
  \BibitemOpen
  \bibfield  {author} {\bibinfo {author} {\bibfnamefont {Y.-C.}\ \bibnamefont
  {Tsai}}\ and\ \bibinfo {author} {\bibfnamefont {K.~A.}\ \bibnamefont
  {Johnson}},\ }\href@noop {} {\bibfield  {journal} {\bibinfo  {journal}
  {Biochemistry}\ }\textbf {\bibinfo {volume} {45}},\ \bibinfo {pages} {9675}
  (\bibinfo {year} {2006})}\BibitemShut {NoStop}%
\bibitem [{\citenamefont {Xie}(2009)}]{xie2009possible}%
  \BibitemOpen
  \bibfield  {author} {\bibinfo {author} {\bibfnamefont {P.}~\bibnamefont
  {Xie}},\ }\href@noop {} {\bibfield  {journal} {\bibinfo  {journal} {Journal
  of theoretical biology}\ }\textbf {\bibinfo {volume} {259}},\ \bibinfo
  {pages} {434} (\bibinfo {year} {2009})}\BibitemShut {NoStop}%
\bibitem [{\citenamefont {Sharma}\ and\ \citenamefont
  {Chowdhury}(2012)}]{sharma2012error}%
  \BibitemOpen
  \bibfield  {author} {\bibinfo {author} {\bibfnamefont {A.~K.}\ \bibnamefont
  {Sharma}}\ and\ \bibinfo {author} {\bibfnamefont {D.}~\bibnamefont
  {Chowdhury}},\ }\href@noop {} {\bibfield  {journal} {\bibinfo  {journal}
  {Physical Review E}\ }\textbf {\bibinfo {volume} {86}},\ \bibinfo {pages}
  {011913} (\bibinfo {year} {2012})}\BibitemShut {NoStop}%
\bibitem [{\citenamefont {Lieberman}\ \emph {et~al.}(2013)\citenamefont
  {Lieberman}, \citenamefont {Dahl}, \citenamefont {Mai}, \citenamefont {Cox},
  \citenamefont {Akeson},\ and\ \citenamefont {Wang}}]{lieberman2013kinetic}%
  \BibitemOpen
  \bibfield  {author} {\bibinfo {author} {\bibfnamefont {K.~R.}\ \bibnamefont
  {Lieberman}}, \bibinfo {author} {\bibfnamefont {J.~M.}\ \bibnamefont {Dahl}},
  \bibinfo {author} {\bibfnamefont {A.~H.}\ \bibnamefont {Mai}}, \bibinfo
  {author} {\bibfnamefont {A.}~\bibnamefont {Cox}}, \bibinfo {author}
  {\bibfnamefont {M.}~\bibnamefont {Akeson}}, \ and\ \bibinfo {author}
  {\bibfnamefont {H.}~\bibnamefont {Wang}},\ }\href@noop {} {\bibfield
  {journal} {\bibinfo  {journal} {Journal of the American Chemical Society}\
  }\textbf {\bibinfo {volume} {135}},\ \bibinfo {pages} {9149} (\bibinfo {year}
  {2013})}\BibitemShut {NoStop}%
\bibitem [{\citenamefont {Lieberman}, \citenamefont {Dahl},\ and\ \citenamefont
  {Wang}(2014)}]{lieberman2014kinetic}%
  \BibitemOpen
  \bibfield  {author} {\bibinfo {author} {\bibfnamefont {K.~R.}\ \bibnamefont
  {Lieberman}}, \bibinfo {author} {\bibfnamefont {J.~M.}\ \bibnamefont {Dahl}},
  \ and\ \bibinfo {author} {\bibfnamefont {H.}~\bibnamefont {Wang}},\
  }\href@noop {} {\bibfield  {journal} {\bibinfo  {journal} {Journal of the
  American Chemical Society}\ }\textbf {\bibinfo {volume} {136}},\ \bibinfo
  {pages} {7117} (\bibinfo {year} {2014})}\BibitemShut {NoStop}%
\bibitem [{\citenamefont {Ollis}\ \emph {et~al.}(1985)\citenamefont {Ollis},
  \citenamefont {Brick}, \citenamefont {Hamlin}, \citenamefont {Xuong},\ and\
  \citenamefont {Steitz}}]{ollis1985structure}%
  \BibitemOpen
  \bibfield  {author} {\bibinfo {author} {\bibfnamefont {D.}~\bibnamefont
  {Ollis}}, \bibinfo {author} {\bibfnamefont {P.}~\bibnamefont {Brick}},
  \bibinfo {author} {\bibfnamefont {R.}~\bibnamefont {Hamlin}}, \bibinfo
  {author} {\bibfnamefont {N.}~\bibnamefont {Xuong}}, \ and\ \bibinfo {author}
  {\bibfnamefont {T.}~\bibnamefont {Steitz}},\ }\href@noop {} {\bibfield
  {journal} {\bibinfo  {journal} {Nature}\ } (\bibinfo {year}
  {1985})}\BibitemShut {NoStop}%
\bibitem [{\citenamefont {Berman}\ \emph {et~al.}(2007)\citenamefont {Berman},
  \citenamefont {Kamtekar}, \citenamefont {Goodman}, \citenamefont
  {L{\'a}zaro}, \citenamefont {de~Vega}, \citenamefont {Blanco}, \citenamefont
  {Salas},\ and\ \citenamefont {Steitz}}]{berman2007structures}%
  \BibitemOpen
  \bibfield  {author} {\bibinfo {author} {\bibfnamefont {A.~J.}\ \bibnamefont
  {Berman}}, \bibinfo {author} {\bibfnamefont {S.}~\bibnamefont {Kamtekar}},
  \bibinfo {author} {\bibfnamefont {J.~L.}\ \bibnamefont {Goodman}}, \bibinfo
  {author} {\bibfnamefont {J.~M.}\ \bibnamefont {L{\'a}zaro}}, \bibinfo
  {author} {\bibfnamefont {M.}~\bibnamefont {de~Vega}}, \bibinfo {author}
  {\bibfnamefont {L.}~\bibnamefont {Blanco}}, \bibinfo {author} {\bibfnamefont
  {M.}~\bibnamefont {Salas}}, \ and\ \bibinfo {author} {\bibfnamefont {T.~A.}\
  \bibnamefont {Steitz}},\ }\href@noop {} {\bibfield  {journal} {\bibinfo
  {journal} {The EMBO journal}\ }\textbf {\bibinfo {volume} {26}},\ \bibinfo
  {pages} {3494} (\bibinfo {year} {2007})}\BibitemShut {NoStop}%
\bibitem [{\citenamefont {Doublie}\ \emph {et~al.}(1998)\citenamefont
  {Doublie}, \citenamefont {Tabor}, \citenamefont {Long}, \citenamefont
  {Richardson},\ and\ \citenamefont {Ellenberger}}]{doublie1998crystal}%
  \BibitemOpen
  \bibfield  {author} {\bibinfo {author} {\bibfnamefont {S.}~\bibnamefont
  {Doublie}}, \bibinfo {author} {\bibfnamefont {S.}~\bibnamefont {Tabor}},
  \bibinfo {author} {\bibfnamefont {A.~M.}\ \bibnamefont {Long}}, \bibinfo
  {author} {\bibfnamefont {C.~C.}\ \bibnamefont {Richardson}}, \ and\ \bibinfo
  {author} {\bibfnamefont {T.}~\bibnamefont {Ellenberger}},\ }\href@noop {}
  {\bibfield  {journal} {\bibinfo  {journal} {nature}\ }\textbf {\bibinfo
  {volume} {391}},\ \bibinfo {pages} {251} (\bibinfo {year}
  {1998})}\BibitemShut {NoStop}%
\bibitem [{\citenamefont {Kamtekar}\ \emph {et~al.}(2004)\citenamefont
  {Kamtekar}, \citenamefont {Berman}, \citenamefont {Wang}, \citenamefont
  {L{\'a}zaro}, \citenamefont {de~Vega}, \citenamefont {Blanco}, \citenamefont
  {Salas},\ and\ \citenamefont {Steitz}}]{kamtekar2004insights}%
  \BibitemOpen
  \bibfield  {author} {\bibinfo {author} {\bibfnamefont {S.}~\bibnamefont
  {Kamtekar}}, \bibinfo {author} {\bibfnamefont {A.~J.}\ \bibnamefont
  {Berman}}, \bibinfo {author} {\bibfnamefont {J.}~\bibnamefont {Wang}},
  \bibinfo {author} {\bibfnamefont {J.~M.}\ \bibnamefont {L{\'a}zaro}},
  \bibinfo {author} {\bibfnamefont {M.}~\bibnamefont {de~Vega}}, \bibinfo
  {author} {\bibfnamefont {L.}~\bibnamefont {Blanco}}, \bibinfo {author}
  {\bibfnamefont {M.}~\bibnamefont {Salas}}, \ and\ \bibinfo {author}
  {\bibfnamefont {T.~A.}\ \bibnamefont {Steitz}},\ }\href@noop {} {\bibfield
  {journal} {\bibinfo  {journal} {Molecular cell}\ }\textbf {\bibinfo {volume}
  {16}},\ \bibinfo {pages} {609} (\bibinfo {year} {2004})}\BibitemShut
  {NoStop}%
\bibitem [{\citenamefont {Wang}\ \emph {et~al.}(1997)\citenamefont {Wang},
  \citenamefont {Sattar}, \citenamefont {Wang}, \citenamefont {Karam},
  \citenamefont {Konigsberg},\ and\ \citenamefont {Steitz}}]{wang1997crystal}%
  \BibitemOpen
  \bibfield  {author} {\bibinfo {author} {\bibfnamefont {J.}~\bibnamefont
  {Wang}}, \bibinfo {author} {\bibfnamefont {A.~A.}\ \bibnamefont {Sattar}},
  \bibinfo {author} {\bibfnamefont {C.}~\bibnamefont {Wang}}, \bibinfo {author}
  {\bibfnamefont {J.}~\bibnamefont {Karam}}, \bibinfo {author} {\bibfnamefont
  {W.}~\bibnamefont {Konigsberg}}, \ and\ \bibinfo {author} {\bibfnamefont
  {T.}~\bibnamefont {Steitz}},\ }\href@noop {} {\bibfield  {journal} {\bibinfo
  {journal} {Cell}\ }\textbf {\bibinfo {volume} {89}},\ \bibinfo {pages} {1087}
  (\bibinfo {year} {1997})}\BibitemShut {NoStop}%
\bibitem [{\citenamefont {Donlin}, \citenamefont {Patel},\ and\ \citenamefont
  {Johnson}(1991)}]{donlin1991kinetic}%
  \BibitemOpen
  \bibfield  {author} {\bibinfo {author} {\bibfnamefont {M.~J.}\ \bibnamefont
  {Donlin}}, \bibinfo {author} {\bibfnamefont {S.~S.}\ \bibnamefont {Patel}}, \
  and\ \bibinfo {author} {\bibfnamefont {K.~A.}\ \bibnamefont {Johnson}},\
  }\href@noop {} {\bibfield  {journal} {\bibinfo  {journal} {Biochemistry}\
  }\textbf {\bibinfo {volume} {30}},\ \bibinfo {pages} {538} (\bibinfo {year}
  {1991})}\BibitemShut {NoStop}%
\bibitem [{\citenamefont {Johnson}\ and\ \citenamefont
  {Johnson}(2001{\natexlab{a}})}]{johnson2001exonuclease}%
  \BibitemOpen
  \bibfield  {author} {\bibinfo {author} {\bibfnamefont {A.~A.}\ \bibnamefont
  {Johnson}}\ and\ \bibinfo {author} {\bibfnamefont {K.~A.}\ \bibnamefont
  {Johnson}},\ }\href@noop {} {\bibfield  {journal} {\bibinfo  {journal}
  {Journal of Biological Chemistry}\ }\textbf {\bibinfo {volume} {276}},\
  \bibinfo {pages} {38097} (\bibinfo {year} {2001}{\natexlab{a}})}\BibitemShut
  {NoStop}%
\bibitem [{\citenamefont {Lamichhane}\ \emph {et~al.}(2013)\citenamefont
  {Lamichhane}, \citenamefont {Berezhna}, \citenamefont {Gill}, \citenamefont
  {Van~der Schans},\ and\ \citenamefont {Millar}}]{lamichhane2013dynamics}%
  \BibitemOpen
  \bibfield  {author} {\bibinfo {author} {\bibfnamefont {R.}~\bibnamefont
  {Lamichhane}}, \bibinfo {author} {\bibfnamefont {S.~Y.}\ \bibnamefont
  {Berezhna}}, \bibinfo {author} {\bibfnamefont {J.~P.}\ \bibnamefont {Gill}},
  \bibinfo {author} {\bibfnamefont {E.}~\bibnamefont {Van~der Schans}}, \ and\
  \bibinfo {author} {\bibfnamefont {D.~P.}\ \bibnamefont {Millar}},\
  }\href@noop {} {\bibfield  {journal} {\bibinfo  {journal} {Journal of the
  American Chemical Society}\ }\textbf {\bibinfo {volume} {135}},\ \bibinfo
  {pages} {4735} (\bibinfo {year} {2013})}\BibitemShut {NoStop}%
\bibitem [{\citenamefont {Beese}, \citenamefont {Derbyshire},\ and\
  \citenamefont {Steitz}(1993)}]{beese1993structure}%
  \BibitemOpen
  \bibfield  {author} {\bibinfo {author} {\bibfnamefont {L.~S.}\ \bibnamefont
  {Beese}}, \bibinfo {author} {\bibfnamefont {V.}~\bibnamefont {Derbyshire}}, \
  and\ \bibinfo {author} {\bibfnamefont {T.~A.}\ \bibnamefont {Steitz}},\
  }\href@noop {} {\bibfield  {journal} {\bibinfo  {journal} {Science}\ }\textbf
  {\bibinfo {volume} {260}},\ \bibinfo {pages} {352} (\bibinfo {year}
  {1993})}\BibitemShut {NoStop}%
\bibitem [{\citenamefont {Shu}\ \emph {et~al.}(2015)\citenamefont {Shu},
  \citenamefont {Song}, \citenamefont {Ou-Yang},\ and\ \citenamefont
  {Li}}]{shu2015general}%
  \BibitemOpen
  \bibfield  {author} {\bibinfo {author} {\bibfnamefont {Y.-G.}\ \bibnamefont
  {Shu}}, \bibinfo {author} {\bibfnamefont {Y.-S.}\ \bibnamefont {Song}},
  \bibinfo {author} {\bibfnamefont {Z.-C.}\ \bibnamefont {Ou-Yang}}, \ and\
  \bibinfo {author} {\bibfnamefont {M.}~\bibnamefont {Li}},\ }\href@noop {}
  {\bibfield  {journal} {\bibinfo  {journal} {Journal of Physics: Condensed
  Matter}\ }\textbf {\bibinfo {volume} {27}},\ \bibinfo {pages} {235105}
  (\bibinfo {year} {2015})}\BibitemShut {NoStop}%
\bibitem [{\citenamefont {Wong}, \citenamefont {Patel},\ and\ \citenamefont
  {Johnson}(1991)}]{wong1991induced}%
  \BibitemOpen
  \bibfield  {author} {\bibinfo {author} {\bibfnamefont {I.}~\bibnamefont
  {Wong}}, \bibinfo {author} {\bibfnamefont {S.~S.}\ \bibnamefont {Patel}}, \
  and\ \bibinfo {author} {\bibfnamefont {K.~A.}\ \bibnamefont {Johnson}},\
  }\href@noop {} {\bibfield  {journal} {\bibinfo  {journal} {Biochemistry}\
  }\textbf {\bibinfo {volume} {30}},\ \bibinfo {pages} {526} (\bibinfo {year}
  {1991})}\BibitemShut {NoStop}%
\bibitem [{\citenamefont {Johnson}\ and\ \citenamefont
  {Johnson}(2001{\natexlab{b}})}]{johnson2001fidelity}%
  \BibitemOpen
  \bibfield  {author} {\bibinfo {author} {\bibfnamefont {A.~A.}\ \bibnamefont
  {Johnson}}\ and\ \bibinfo {author} {\bibfnamefont {K.~A.}\ \bibnamefont
  {Johnson}},\ }\href@noop {} {\bibfield  {journal} {\bibinfo  {journal}
  {Journal of Biological Chemistry}\ }\textbf {\bibinfo {volume} {276}},\
  \bibinfo {pages} {38090} (\bibinfo {year} {2001}{\natexlab{b}})}\BibitemShut
  {NoStop}%
\bibitem [{\citenamefont {Gillespie}(1977)}]{gillespie1977exact}%
  \BibitemOpen
  \bibfield  {author} {\bibinfo {author} {\bibfnamefont {D.~T.}\ \bibnamefont
  {Gillespie}},\ }\href@noop {} {\bibfield  {journal} {\bibinfo  {journal} {The
  journal of physical chemistry}\ }\textbf {\bibinfo {volume} {81}},\ \bibinfo
  {pages} {2340} (\bibinfo {year} {1977})}\BibitemShut {NoStop}%
\bibitem [{\citenamefont {Johnson}\ and\ \citenamefont
  {Beese}(2004)}]{johnson2004structures}%
  \BibitemOpen
  \bibfield  {author} {\bibinfo {author} {\bibfnamefont {S.~J.}\ \bibnamefont
  {Johnson}}\ and\ \bibinfo {author} {\bibfnamefont {L.~S.}\ \bibnamefont
  {Beese}},\ }\href@noop {} {\bibfield  {journal} {\bibinfo  {journal} {Cell}\
  }\textbf {\bibinfo {volume} {116}},\ \bibinfo {pages} {803} (\bibinfo {year}
  {2004})}\BibitemShut {NoStop}%
\bibitem [{\citenamefont {Tindall}\ and\ \citenamefont
  {Kunkel}(1988)}]{tindall1988fidelity}%
  \BibitemOpen
  \bibfield  {author} {\bibinfo {author} {\bibfnamefont {K.~R.}\ \bibnamefont
  {Tindall}}\ and\ \bibinfo {author} {\bibfnamefont {T.~A.}\ \bibnamefont
  {Kunkel}},\ }\href@noop {} {\bibfield  {journal} {\bibinfo  {journal}
  {Biochemistry}\ }\textbf {\bibinfo {volume} {27}},\ \bibinfo {pages} {6008}
  (\bibinfo {year} {1988})}\BibitemShut {NoStop}%
\bibitem [{\citenamefont {Lundberg}\ \emph {et~al.}(1991)\citenamefont
  {Lundberg}, \citenamefont {Dan}, \citenamefont {Adams}, \citenamefont
  {Short}, \citenamefont {Sorge},\ and\ \citenamefont
  {Mathur}}]{Lundberg1991High}%
  \BibitemOpen
  \bibfield  {author} {\bibinfo {author} {\bibfnamefont {K.~S.}\ \bibnamefont
  {Lundberg}}, \bibinfo {author} {\bibfnamefont {D.~S.}\ \bibnamefont {Dan}},
  \bibinfo {author} {\bibfnamefont {M.~W.~W.}\ \bibnamefont {Adams}}, \bibinfo
  {author} {\bibfnamefont {J.~M.}\ \bibnamefont {Short}}, \bibinfo {author}
  {\bibfnamefont {J.~A.}\ \bibnamefont {Sorge}}, \ and\ \bibinfo {author}
  {\bibfnamefont {E.~J.}\ \bibnamefont {Mathur}},\ }\href@noop {} {\bibfield
  {journal} {\bibinfo  {journal} {Gene}\ }\textbf {\bibinfo {volume} {108}},\
  \bibinfo {pages} {1} (\bibinfo {year} {1991})}\BibitemShut {NoStop}%
\bibitem [{\citenamefont {Kokoska}\ \emph {et~al.}(2002)\citenamefont
  {Kokoska}, \citenamefont {Bebenek}, \citenamefont {Boudsocq}, \citenamefont
  {Woodgate},\ and\ \citenamefont {Kunkel}}]{kokoska2002low}%
  \BibitemOpen
  \bibfield  {author} {\bibinfo {author} {\bibfnamefont {R.~J.}\ \bibnamefont
  {Kokoska}}, \bibinfo {author} {\bibfnamefont {K.}~\bibnamefont {Bebenek}},
  \bibinfo {author} {\bibfnamefont {F.}~\bibnamefont {Boudsocq}}, \bibinfo
  {author} {\bibfnamefont {R.}~\bibnamefont {Woodgate}}, \ and\ \bibinfo
  {author} {\bibfnamefont {T.~A.}\ \bibnamefont {Kunkel}},\ }\href@noop {}
  {\bibfield  {journal} {\bibinfo  {journal} {Journal of Biological Chemistry}\
  }\textbf {\bibinfo {volume} {277}},\ \bibinfo {pages} {19633} (\bibinfo
  {year} {2002})}\BibitemShut {NoStop}%
\bibitem [{\citenamefont {Boosalis}, \citenamefont {Petruska},\ and\
  \citenamefont {Goodman}(1987)}]{boosalis1987dna}%
  \BibitemOpen
  \bibfield  {author} {\bibinfo {author} {\bibfnamefont {M.~S.}\ \bibnamefont
  {Boosalis}}, \bibinfo {author} {\bibfnamefont {J.}~\bibnamefont {Petruska}},
  \ and\ \bibinfo {author} {\bibfnamefont {M.}~\bibnamefont {Goodman}},\
  }\href@noop {} {\bibfield  {journal} {\bibinfo  {journal} {Journal of
  Biological Chemistry}\ }\textbf {\bibinfo {volume} {262}},\ \bibinfo {pages}
  {14689} (\bibinfo {year} {1987})}\BibitemShut {NoStop}%
\bibitem [{\citenamefont {Goodman}\ \emph {et~al.}(1993)\citenamefont
  {Goodman}, \citenamefont {Creighton}, \citenamefont {Bloom}, \citenamefont
  {Petruska},\ and\ \citenamefont {Kunkel}}]{goodman1993biochemical}%
  \BibitemOpen
  \bibfield  {author} {\bibinfo {author} {\bibfnamefont {M.~F.}\ \bibnamefont
  {Goodman}}, \bibinfo {author} {\bibfnamefont {S.}~\bibnamefont {Creighton}},
  \bibinfo {author} {\bibfnamefont {L.~B.}\ \bibnamefont {Bloom}}, \bibinfo
  {author} {\bibfnamefont {J.}~\bibnamefont {Petruska}}, \ and\ \bibinfo
  {author} {\bibfnamefont {T.~A.}\ \bibnamefont {Kunkel}},\ }\href@noop {}
  {\bibfield  {journal} {\bibinfo  {journal} {Critical reviews in biochemistry
  and molecular biology}\ }\textbf {\bibinfo {volume} {28}},\ \bibinfo {pages}
  {83} (\bibinfo {year} {1993})}\BibitemShut {NoStop}%
\bibitem [{\citenamefont {Bebenek}\ \emph {et~al.}(1990)\citenamefont
  {Bebenek}, \citenamefont {Joyce}, \citenamefont {Fitzgerald},\ and\
  \citenamefont {Kunkel}}]{bebenek1990fidelity}%
  \BibitemOpen
  \bibfield  {author} {\bibinfo {author} {\bibfnamefont {K.}~\bibnamefont
  {Bebenek}}, \bibinfo {author} {\bibfnamefont {C.}~\bibnamefont {Joyce}},
  \bibinfo {author} {\bibfnamefont {M.~P.}\ \bibnamefont {Fitzgerald}}, \ and\
  \bibinfo {author} {\bibfnamefont {T.}~\bibnamefont {Kunkel}},\ }\href@noop {}
  {\bibfield  {journal} {\bibinfo  {journal} {Journal of Biological Chemistry}\
  }\textbf {\bibinfo {volume} {265}},\ \bibinfo {pages} {13878} (\bibinfo
  {year} {1990})}\BibitemShut {NoStop}%
\bibitem [{\citenamefont {Kuchta}\ \emph {et~al.}(1987)\citenamefont {Kuchta},
  \citenamefont {Mizrahi}, \citenamefont {Benkovic}, \citenamefont {Johnson},\
  and\ \citenamefont {Benkovic}}]{kuchta1987kinetic}%
  \BibitemOpen
  \bibfield  {author} {\bibinfo {author} {\bibfnamefont {R.}~\bibnamefont
  {Kuchta}}, \bibinfo {author} {\bibfnamefont {V.}~\bibnamefont {Mizrahi}},
  \bibinfo {author} {\bibfnamefont {P.}~\bibnamefont {Benkovic}}, \bibinfo
  {author} {\bibfnamefont {K.}~\bibnamefont {Johnson}}, \ and\ \bibinfo
  {author} {\bibfnamefont {S.}~\bibnamefont {Benkovic}},\ }\href@noop {}
  {\bibfield  {journal} {\bibinfo  {journal} {Biochemistry}\ }\textbf {\bibinfo
  {volume} {26}},\ \bibinfo {pages} {8410} (\bibinfo {year}
  {1987})}\BibitemShut {NoStop}%
\bibitem [{\citenamefont {Kuchta}, \citenamefont {Benkovic},\ and\
  \citenamefont {Benkovic}(1988)}]{kuchta1988kinetic}%
  \BibitemOpen
  \bibfield  {author} {\bibinfo {author} {\bibfnamefont {R.~D.}\ \bibnamefont
  {Kuchta}}, \bibinfo {author} {\bibfnamefont {P.}~\bibnamefont {Benkovic}}, \
  and\ \bibinfo {author} {\bibfnamefont {S.~J.}\ \bibnamefont {Benkovic}},\
  }\href@noop {} {\bibfield  {journal} {\bibinfo  {journal} {Biochemistry}\
  }\textbf {\bibinfo {volume} {27}},\ \bibinfo {pages} {6716} (\bibinfo {year}
  {1988})}\BibitemShut {NoStop}%
\bibitem [{\citenamefont {Fiala}\ and\ \citenamefont
  {Suo}(2004)}]{fiala2004pre}%
  \BibitemOpen
  \bibfield  {author} {\bibinfo {author} {\bibfnamefont {K.~A.}\ \bibnamefont
  {Fiala}}\ and\ \bibinfo {author} {\bibfnamefont {Z.}~\bibnamefont {Suo}},\
  }\href@noop {} {\bibfield  {journal} {\bibinfo  {journal} {Biochemistry}\
  }\textbf {\bibinfo {volume} {43}},\ \bibinfo {pages} {2106} (\bibinfo {year}
  {2004})}\BibitemShut {NoStop}%
\bibitem [{\citenamefont {Boudsocq}\ \emph {et~al.}(2001)\citenamefont
  {Boudsocq}, \citenamefont {Iwai}, \citenamefont {Hanaoka},\ and\
  \citenamefont {Woodgate}}]{boudsocq2001sulfolobus}%
  \BibitemOpen
  \bibfield  {author} {\bibinfo {author} {\bibfnamefont {F.}~\bibnamefont
  {Boudsocq}}, \bibinfo {author} {\bibfnamefont {S.}~\bibnamefont {Iwai}},
  \bibinfo {author} {\bibfnamefont {F.}~\bibnamefont {Hanaoka}}, \ and\
  \bibinfo {author} {\bibfnamefont {R.}~\bibnamefont {Woodgate}},\ }\href@noop
  {} {\bibfield  {journal} {\bibinfo  {journal} {Nucleic Acids Research}\
  }\textbf {\bibinfo {volume} {29}},\ \bibinfo {pages} {4607} (\bibinfo {year}
  {2001})}\BibitemShut {NoStop}%
\bibitem [{\citenamefont {Kunkel}\ \emph {et~al.}(1992)\citenamefont {Kunkel},
  \citenamefont {Hizi}, \citenamefont {Shaharabany}, \citenamefont {Tsygankov},
  \citenamefont {Br{\"o}ker}, \citenamefont {Fargnoli}, \citenamefont
  {Ledbetter}, \citenamefont {Bolen}, \citenamefont {De~Vivo}, \citenamefont
  {Chen} \emph {et~al.}}]{kunkel1992dna}%
  \BibitemOpen
  \bibfield  {author} {\bibinfo {author} {\bibfnamefont {T.}~\bibnamefont
  {Kunkel}}, \bibinfo {author} {\bibfnamefont {A.}~\bibnamefont {Hizi}},
  \bibinfo {author} {\bibfnamefont {M.}~\bibnamefont {Shaharabany}}, \bibinfo
  {author} {\bibfnamefont {A.}~\bibnamefont {Tsygankov}}, \bibinfo {author}
  {\bibfnamefont {B.}~\bibnamefont {Br{\"o}ker}}, \bibinfo {author}
  {\bibfnamefont {J.}~\bibnamefont {Fargnoli}}, \bibinfo {author}
  {\bibfnamefont {J.}~\bibnamefont {Ledbetter}}, \bibinfo {author}
  {\bibfnamefont {J.}~\bibnamefont {Bolen}}, \bibinfo {author} {\bibfnamefont
  {M.}~\bibnamefont {De~Vivo}}, \bibinfo {author} {\bibfnamefont
  {J.}~\bibnamefont {Chen}},  \emph {et~al.},\ }\href@noop {} {\bibfield
  {journal} {\bibinfo  {journal} {J. Biol. Chem}\ }\textbf {\bibinfo {volume}
  {267}} (\bibinfo {year} {1992})}\BibitemShut {NoStop}%
\bibitem [{\citenamefont {Cady}\ and\ \citenamefont
  {Qian}(2009)}]{cady2009open}%
  \BibitemOpen
  \bibfield  {author} {\bibinfo {author} {\bibfnamefont {F.}~\bibnamefont
  {Cady}}\ and\ \bibinfo {author} {\bibfnamefont {H.}~\bibnamefont {Qian}},\
  }\href@noop {} {\bibfield  {journal} {\bibinfo  {journal} {Physical biology}\
  }\textbf {\bibinfo {volume} {6}},\ \bibinfo {pages} {036011} (\bibinfo {year}
  {2009})}\BibitemShut {NoStop}%
\bibitem [{\citenamefont {Bloom}\ \emph {et~al.}(1997)\citenamefont {Bloom},
  \citenamefont {Chen}, \citenamefont {Fygenson}, \citenamefont {Turner},
  \citenamefont {O’Donnell},\ and\ \citenamefont
  {Goodman}}]{bloom1997fidelity}%
  \BibitemOpen
  \bibfield  {author} {\bibinfo {author} {\bibfnamefont {L.~B.}\ \bibnamefont
  {Bloom}}, \bibinfo {author} {\bibfnamefont {X.}~\bibnamefont {Chen}},
  \bibinfo {author} {\bibfnamefont {D.~K.}\ \bibnamefont {Fygenson}}, \bibinfo
  {author} {\bibfnamefont {J.}~\bibnamefont {Turner}}, \bibinfo {author}
  {\bibfnamefont {M.}~\bibnamefont {O’Donnell}}, \ and\ \bibinfo {author}
  {\bibfnamefont {M.~F.}\ \bibnamefont {Goodman}},\ }\href@noop {} {\bibfield
  {journal} {\bibinfo  {journal} {Journal of Biological Chemistry}\ }\textbf
  {\bibinfo {volume} {272}},\ \bibinfo {pages} {27919} (\bibinfo {year}
  {1997})}\BibitemShut {NoStop}%
\bibitem [{\citenamefont {Miller}\ and\ \citenamefont
  {Perrino}(1996)}]{miller1996kinetic}%
  \BibitemOpen
  \bibfield  {author} {\bibinfo {author} {\bibfnamefont {H.}~\bibnamefont
  {Miller}}\ and\ \bibinfo {author} {\bibfnamefont {F.~W.}\ \bibnamefont
  {Perrino}},\ }\href@noop {} {\bibfield  {journal} {\bibinfo  {journal}
  {Biochemistry}\ }\textbf {\bibinfo {volume} {35}},\ \bibinfo {pages} {12919}
  (\bibinfo {year} {1996})}\BibitemShut {NoStop}%
\bibitem [{\citenamefont {Capson}\ \emph {et~al.}(1992)\citenamefont {Capson},
  \citenamefont {Peliska}, \citenamefont {Kaboord}, \citenamefont {Frey},
  \citenamefont {Lively}, \citenamefont {Dahlberg},\ and\ \citenamefont
  {Benkovic}}]{capson1992kinetic}%
  \BibitemOpen
  \bibfield  {author} {\bibinfo {author} {\bibfnamefont {T.~L.}\ \bibnamefont
  {Capson}}, \bibinfo {author} {\bibfnamefont {J.~A.}\ \bibnamefont {Peliska}},
  \bibinfo {author} {\bibfnamefont {B.~F.}\ \bibnamefont {Kaboord}}, \bibinfo
  {author} {\bibfnamefont {M.~W.}\ \bibnamefont {Frey}}, \bibinfo {author}
  {\bibfnamefont {C.}~\bibnamefont {Lively}}, \bibinfo {author} {\bibfnamefont
  {M.}~\bibnamefont {Dahlberg}}, \ and\ \bibinfo {author} {\bibfnamefont
  {S.~J.}\ \bibnamefont {Benkovic}},\ }\href@noop {} {\bibfield  {journal}
  {\bibinfo  {journal} {Biochemistry}\ }\textbf {\bibinfo {volume} {31}},\
  \bibinfo {pages} {10984} (\bibinfo {year} {1992})}\BibitemShut {NoStop}%
\bibitem [{\citenamefont {Esteban}, \citenamefont {Salas},\ and\ \citenamefont
  {Blanco}(1993)}]{esteban1993fidelity}%
  \BibitemOpen
  \bibfield  {author} {\bibinfo {author} {\bibfnamefont {J.~A.}\ \bibnamefont
  {Esteban}}, \bibinfo {author} {\bibfnamefont {M.}~\bibnamefont {Salas}}, \
  and\ \bibinfo {author} {\bibfnamefont {L.}~\bibnamefont {Blanco}},\
  }\href@noop {} {\bibfield  {journal} {\bibinfo  {journal} {Journal of
  Biological Chemistry}\ }\textbf {\bibinfo {volume} {268}},\ \bibinfo {pages}
  {2719} (\bibinfo {year} {1993})}\BibitemShut {NoStop}%
\bibitem [{\citenamefont {Esteban}\ \emph {et~al.}(1994)\citenamefont
  {Esteban}, \citenamefont {Soengas}, \citenamefont {Salas},\ and\
  \citenamefont {Blanco}}]{esteban19943}%
  \BibitemOpen
  \bibfield  {author} {\bibinfo {author} {\bibfnamefont {J.~A.}\ \bibnamefont
  {Esteban}}, \bibinfo {author} {\bibfnamefont {M.~S.}\ \bibnamefont
  {Soengas}}, \bibinfo {author} {\bibfnamefont {M.}~\bibnamefont {Salas}}, \
  and\ \bibinfo {author} {\bibfnamefont {L.}~\bibnamefont {Blanco}},\
  }\href@noop {} {\bibfield  {journal} {\bibinfo  {journal} {Journal of
  Biological Chemistry}\ }\textbf {\bibinfo {volume} {269}},\ \bibinfo {pages}
  {31946} (\bibinfo {year} {1994})}\BibitemShut {NoStop}%
\bibitem [{\citenamefont {Petruska}\ \emph {et~al.}(1988)\citenamefont
  {Petruska}, \citenamefont {Goodman}, \citenamefont {Boosalis}, \citenamefont
  {Sowers}, \citenamefont {Cheong},\ and\ \citenamefont
  {Tinoco}}]{petruska1988comparison}%
  \BibitemOpen
  \bibfield  {author} {\bibinfo {author} {\bibfnamefont {J.}~\bibnamefont
  {Petruska}}, \bibinfo {author} {\bibfnamefont {M.~F.}\ \bibnamefont
  {Goodman}}, \bibinfo {author} {\bibfnamefont {M.~S.}\ \bibnamefont
  {Boosalis}}, \bibinfo {author} {\bibfnamefont {L.~C.}\ \bibnamefont
  {Sowers}}, \bibinfo {author} {\bibfnamefont {C.}~\bibnamefont {Cheong}}, \
  and\ \bibinfo {author} {\bibfnamefont {I.}~\bibnamefont {Tinoco}},\
  }\href@noop {} {\bibfield  {journal} {\bibinfo  {journal} {Proceedings of the
  National Academy of Sciences}\ }\textbf {\bibinfo {volume} {85}},\ \bibinfo
  {pages} {6252} (\bibinfo {year} {1988})}\BibitemShut {NoStop}%
\bibitem [{\citenamefont {Pierre~Gaspard}(2014)}]{gaspard2014kinetics}%
  \BibitemOpen
  \bibfield  {author} {\bibinfo {author} {\bibfnamefont {D.~A.}\ \bibnamefont
  {Pierre~Gaspard}},\ }\href@noop {} {\bibfield  {journal} {\bibinfo  {journal}
  {Journal of Chemical Physics}\ }\textbf {\bibinfo {volume} {141}} (\bibinfo
  {year} {2014})}\BibitemShut {NoStop}%
\end{thebibliography}
\end{document}